\documentclass[aps, prl, reprint, superscriptaddress, amsmath, amssymb]{revtex4-2}
\usepackage{graphicx}
\usepackage{dcolumn}
\usepackage{bm}
\usepackage{color}
\usepackage{braket}
\usepackage{minitoc} 
\begin{document}
\title{
Optical transitions of a single nodal ring in SrAs$_3$: radially and axially resolved characterization
} 

\author{Jiwon Jeon}
 \thanks{These authors contributed equally to this work.}
\affiliation{%
 Natural Science Research Institute, University of Seoul, Seoul 02504, Korea
 }%
 \affiliation{%
 Physics Department, University of Seoul, Seoul 02504, Korea
 }%

\author{Jiho Jang}%
 \thanks{These authors contributed equally to this work.}
 \affiliation{%
 Department of Physics and Astronomy, Seoul National University, Seoul 08826, Korea
 }%

\author{Hoil Kim}%
 \affiliation{%
 Center for Artificial Low Dimensional Electronic Systems, Institute for Basic Science (IBS), Pohang 37673, Korea
 }%
  \affiliation{%
 Department of Physics, Pohang University of Science and Technology (POSTECH), Pohang 37673, Korea
 }%

\author{Taesu Park}%
 \affiliation{%
 Department of Chemistry, Pohang University of Science and Technology (POSTECH), Pohang 37673, Korea
 }%

 \author{DongWook Kim}%
 \affiliation{%
 Department of Physics, Hanyang University, Seoul 04763, Korea
 }%

 \author{Soonjae Moon}%
 \affiliation{%
 Department of Physics, Hanyang University, Seoul 04763, Korea
 }%

\author{Jun Sung Kim}%
 \affiliation{%
 Center for Artificial Low Dimensional Electronic Systems, Institute for Basic Science (IBS), Pohang 37673, Korea
 }%
  \affiliation{%
 Department of Physics, Pohang University of Science and Technology (POSTECH), Pohang 37673, Korea
 }%

\author{Ji Hoon Shim}%
 \affiliation{%
 Department of Chemistry, Pohang University of Science and Technology (POSTECH), Pohang 37673, Korea
 }%

\author{Hongki Min}%
 \email{hmin@snu.ac.kr}
 \affiliation{%
 Department of Physics and Astronomy, Seoul National University, Seoul 08826, Korea
 }%

\author{Eunjip Choi}%
\email{echoi@uos.ac.kr}
 \affiliation{%
 Physics Department, University of Seoul, Seoul 02504, Korea
 }%
 
\date{\today}

\begin{abstract}
SrAs$_3$ is a unique nodal-line semimetal that contains only a single nodal ring in the Brillouin zone, uninterrupted by any trivial bands near the Fermi energy.
We performed axis-resolved optical reflection measurements on SrAs$_3$ and observed that the optical conductivity exhibits flat absorption up to 129 meV in both the radial and axial directions, confirming the robustness of the universal power-law behavior of the nodal ring. 
Furthermore, in conjunction with model and first-principles calculations, the axis-resolved optical conductivity unveiled fundamental properties beyond the flat absorption, including the overlap energy of the topological bands, the spin-orbit coupling gap along the nodal ring, and the geometric properties of the nodal ring such as the average ring radius, ring ellipticity, and velocity anisotropy.
In addition, our temperature-dependent measurements revealed a spectral weight transfer between intraband and interband transitions, indicating a possible violation of the optical sum rule within the measured energy range.
\end{abstract}

\maketitle

Topological semimetals, a class of novel quantum materials characterized by symmetry-protected band contact or crossing, have received significant attention in recent years due to their unique fundamental properties and potential for device applications \cite{RevModPhys.93.025002, RevModPhys.90.015001}. 
Nodal-line semimetals (NLSMs) are a type of topological semimetals that have two energy bands crossing over a continuous line in the $k$-space \cite{PhysRevB.84.235126, PhysRevB.90.115111, PhysRevB.92.081201, PhysRevB.93.155140, Fang_2016}.
The exotic quasiparticle excitations and rich topological properties of NLSMs, including the non-trivial Berry phase around the nodal line, give rise to various intriguing phenomena such as anomalous transverse current \cite{PhysRevB.97.161113}, weak localization \cite{PhysRevLett.122.196603}, and correlation effect \cite{Pezzini2018, Shao2020}.

Optical conductivity measurements serve as a powerful tool in the study of NLSMs for probing the intricate low-energy electronic band structures of materials since they provide direct access to states close to the Fermi energy $E_{\mathrm{F}}$.
In NLSMs, the interband transition between two bands forming the nodal line is predicted to exhibit a distinct power-law behavior of the real part of the optical conductivity, $\sigma_{1}(\omega) \sim \omega^{(d-2)/z}$ with the effective dimension $d=2$, which leads to the flat optical conductivity regardless of the band dispersion power $z$ \cite{PhysRevB.87.125425, Carbotte_2017, PhysRevB.95.214203, PhysRevB.96.155150, PhysRevLett.119.147402}.

However, experimentally, most optical studies of NLSMs face challenges that hinder an in-depth investigation.
Firstly, numerous real materials exhibit multiple nodal lines instead of a single one in the Brillouin zone (BZ).
These nodal lines take the form of open nodal lines \cite{PhysRevLett.117.016602, PhysRevLett.119.187401, Shao2020}, rectangular- \cite{PhysRevB.104.L201115, PhysRevB.107.045115, doi:10.1073/pnas.1809631115} or irregular-shaped \cite{PhysRevB.103.125131, PhysRevB.106.245145} cages, which often become entangled into  complicated manifolds, adding complexity to the analysis.
Secondly, in most real materials, topologically trivial bands coexist at the $E_{\mathrm{F}}$ alongside the nodal line.
These trivial bands create their own intra- and interband transitions that seriously obscure the optical signal from the non-trivial band \cite{PhysRevB.104.L201115, doi:10.1073/pnas.1809631115, C9TC03464A, PhysRevB.93.121113, Feng2017, PhysRevB.103.125131}, making it difficult to investigate the genuine properties of the nodal line. 
Due to these two challenges, the majority of optical experiments  have been limited to demonstrating the presence of the flat conductivity \cite{PhysRevLett.119.187401, Shao2020, PhysRevB.104.L201115, PhysRevB.100.125136, PhysRevB.107.045115, doi:10.1073/pnas.1809631115, PhysRevB.106.075143}. 
To advance the study of NLSMs beyond the flat conductivity, real materials with a simple nodal line and no trivial bands are highly desired.

In this work, we perform optical measurements on $\mathrm{SrAs_3}$, a rare example that features a single, isolated (doubly degenerate) nodal ring in the BZ \cite{PhysRevB.95.045136, PhysRevLett.124.056402, Hosen2020, Kim2022}.
Furthermore, remarkably, it possesses no interrupting trivial bands near $E_{\mathrm{F}}$  but consists of only the non-trivial topological bands \cite{PhysRevB.95.045136, Hosen2020, https://doi.org/10.48550/arxiv.2211.05978}.
These outstanding properties of $\mathrm{SrAs_3}$ provide a unique opportunity for an in-depth investigation beyond the power-law behavior.
Optical conductivities for both radial and axial directions of the nodal ring are measured across a various temperature range.
The axis-resolved optical conductivity is combined with model and first-principles calculations, unveiling fundamental properties such as the overlap energy of the non-trivial bands, the spin-orbit coupling (SOC) that opens a gap along the nodal ring, the axis-dependent velocity anisotropy along with the average ring radius and ring ellipticity.
In addition, our temperature-dependent measurements reveal an anomalous spectral weight transfer between the intraband and interband transitions.

A single crystal $\rm{SrAs_3}$ was synthesized and thoroughly characterized \cite{Kim2022}.
An axis-resolved optical measurement was performed using polarized light.
The experimental details are described in the Supplemental Material (SM) Sec. I \cite{SM}.

\nocite{Kim2022, Homes:93, doi:10.1063/1.5143061, PhysRevLett.102.226401, togo2015first, PhysRevB.54.11169, PhysRevB.95.045136, PhysRevLett.118.176402, mahan2000many, PhysRevLett.119.147402, doi:10.1073/pnas.1809631115}

\begin{figure}
\includegraphics[width=1\columnwidth]{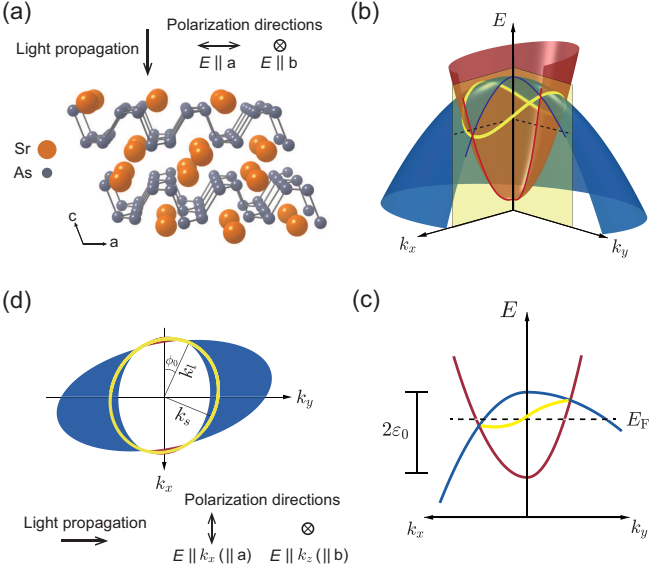}
\caption{\label{fig:fig1}
Schematic drawing of (a) crystal structure and (b) band structure of $\rm{SrAs_3}$.
(c) A simplified view of (b) where $2\varepsilon_0$ and $E_{\mathrm{F}}$ are the band overlap energy and Fermi energy, respectively.
(d) The nodal ring is elliptically elongated where $k_{l}$, $k_{s}$, and $\phi_{0}$ are the lengths of the semi-major and semi-minor axes and the rotation angle, respectively.
The blue and red pockets represent the hole and electron carriers, respectively.
The yellow curves in (b-d) highlight the nodal ring.
The SOC gap is omitted in the figures for clarity.
In the optical measurement, incident light is polarized as specified in (a) and (d).}
\end{figure}

Figure 1 presents the crystal structure and electronic band structure of $\rm{SrAs_3}$.
The lattice of $\rm{SrAs_3}$, which belongs to the $C2/m$ space group, consists of As-planes separated by Sr atoms that are stacked along the $c$-axis.
Figure 1(b) displays the conduction and valence bands, which are both mainly from As $4p$ orbitals, that intersect each other leading to the formation of a nodal ring.
Figure 1(c), a simplified view of Fig. 1(b), shows that the two bands overlap by $2\varepsilon_0$. They disperse asymmetrically along the $k_x$ and $k_y$ directions, causing the nodal ring to tilt.
Figure 1(d) displays the nodal ring in $k$-space.
The ring is elliptically elongated and rotated on the $k_x$-$k_y$ plane, as shown by ARPES and DFT studies \cite{PhysRevLett.124.056402, PhysRevB.95.045136}.
The two light polarizations, $E \parallel k_x$ and $E \parallel k_z$, employed in our measurement probe the radial and axial directions of the nodal ring, respectively. (Here we refer to $E \parallel k_x$ as the radial direction,  even though the nodal ring is a rotated ellipse.) 
In $\rm{SrAs_3}$, the SOC creates an energy gap along the nodal ring, which is intentionally omitted for clarity in Fig. 1.
The Fermi energy and carrier density of our sample were precisely determined using quantum oscillation measurements as described in Ref. \cite{Kim2022}.
The $E_{\mathrm{F}}$ indicated in Fig. 1(c) shows that the hole is dominant over the electron, where the actual carrier densities are  $n_h$ = 4.2$\times 10^{17}$ cm$^{-3}$ for the hole and $n_e$ = 6.9$\times 10^{15}$ cm$^{-3}$ for the electron, measured from quantum transport \cite{Kim2022} at $T$ = 5 K, respectively, which is schematically represented in the sizes of the blue (hole) and red (electron) pockets as well in Fig. 1(d).

\begin{figure}
\includegraphics[width=1\columnwidth]{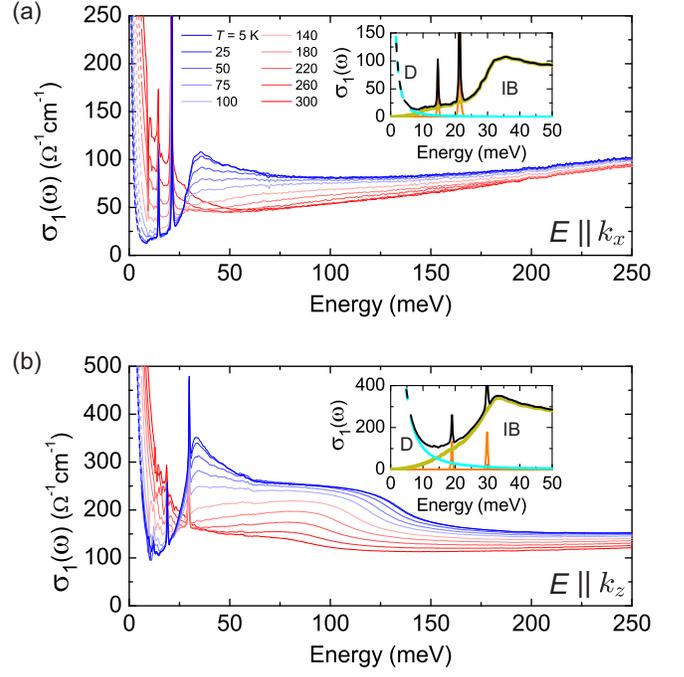} 
\caption{\label{fig:fig2}
Optical conductivity of $\rm{SrAs_3}$ for the two polarization directions (a) $E$ $\parallel$ $k_x$ and (b) $E$ $\parallel$ $k_z$ over 5 K $\leq T \leq$ 300 K. Insets show that $\sigma_{1}(\omega)$ is decomposed into Drude (D), phonon, and interband (IB) contributions. For the range below 6.2 meV,  $\sigma_{1}(\omega)$  is an extrapolation derived from the constrained Kramers-Kronig analysis and is represented as dashed lines.
}
\end{figure}

Figure \ref{fig:fig2} shows the optical conductivity $\sigma_{1}(\omega)$ for $E \parallel k_x$ and $E \parallel k_z$ for 5 K $\leq T \leq$ 300 K. They were obtained by fitting the reflectivity data $R(\omega)$ using the Kramers-Kronig constrained variational dielectric functions of RefFit \cite{doi:10.1063/1.1979470}. $\sigma_{1}(\omega)$ shows a good agreement with the transport conductivity at $\omega$ = 0 (see Fig. S3 in SM Sec. II \cite{SM}).
The raw $R(\omega)$ data are presented in SM Fig. S1 \cite{SM}.  
$\sigma_{1}(\omega)$ consists of a Drude peak at low energy and an interband transition in the high energy range, respectively. In addition, two sharp optical phonon peaks exist between
the Drude and interband conductivities. 
For quantitative analysis,  we decompose $\sigma_{1}(\omega)$ into three components through data fitting: Drude, phonon, and interband (IB) conductivities as $\sigma_1(\omega) = \sigma_{1}^{\mathrm{D}}(\omega) + \sigma_{1}^{\mathrm{Ph}}(\omega) + \sigma_{1}^{\mathrm{IB}}(\omega)$. The details of the fitting analysis are presented in SM Sec. II \cite{SM}.

\begin{figure}
\includegraphics[width=1\columnwidth]{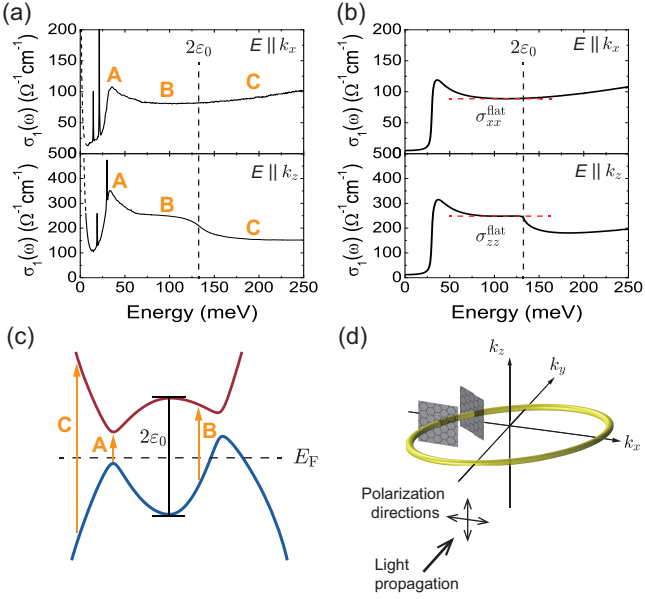}
\caption{\label{fig:fig3}
Optical conductivity obtained from (a) reflectivity measurement at $T$ = 5 K and (b) theoretical calculation for interband transitions at $T=0$ K.
The A, B, and C represent the SOC-induced optical peak, the flat conductivity region, and the interband transition above the band overlap energy, respectively.
The $\sigma^{\rm{flat}}_{xx}$ and $\sigma^{\rm{flat}}_{zz}$ marked by red dashed lines in panel (b) show  the flat conductivity levels.
(c) Schematic band structure near the Fermi energy $E_{\mathrm{F}}$ along the radial direction, where SOC opens a gap of 2$\Delta_{\mathrm{SOC}}$ along the nodal ring. The black solid line indicates the band overlap energy 2$\varepsilon_0$.
The arrows A, B, and C represent the corresponding interband transitions in (a). 
(d) Schematic diagram illustrating the nodal ring as a collection of gapped anisotropic graphene sheets.
}
\end{figure}

In  Fig. 3(a), we focus on the low temperature $\sigma_{1}(\omega)$ result taken at $T = 5$ K.  The Drude peak is described by $\sigma_{1}^{\mathrm{D}}(\omega) = \frac{\omega^2_{p}}{4\pi}\frac{\gamma}{(\gamma^2+\omega^2)}$, where $\omega_{p}$ and $\gamma$ = 1/$\tau$ are the plasma frequency and the carrier scattering rate, respectively.
$\sigma_{1}^{\mathrm{D}}(\omega)$ arises overwhelmingly from the hole carrier
due to its dominance over the electron.
Using the fitting results for $\omega_{p,x}^{2}$ ($E \parallel k_x$) and $\omega_{p,z}^{2}$ ($E \parallel k_z$), summarized in SM Table S1 \cite{SM}, we obtain $\omega_{p,z}^{2}$/$\omega_{p,x}^{2}$ = 8.7 which, from the definition of $\omega_{p}^{2}$ = $\frac{4\pi ne^{2}}{m^*}$ ($n$ = hole density,  $m^{*}$ = effective mass), is equivalent to  $m^*_{x}$/$m^*_{z}$.
It shows that the effective mass of the hole is substantially lighter along the Sr-chain ($b$-axis) than along the Sr-zigzag direction ($a$-axis), similar to black phosphorus \cite{19833544}.
By inserting the hole density $n_{h}$ obtained from transport measurement into $\omega_{p}^{2}$, $m^*_z$ = 0.017$m_0$ and  $m^*_x$ = 0.148$m_0$ are obtained. Furthermore, 
by using $\tau$ = 1/$\gamma$ and $m^{*}$, we calculate the carrier mobility $\mu = \frac{e\tau}{{m^*}}$ to be  
$\mu_{x}$ = 3.9$\times 10^{3}$ $\rm{cm}^2 \thinspace V^{-1} \thinspace s^{-1}$ and $\mu_{z}$ = 2.1$\times 10^{4}$ $\rm{cm}^2 \thinspace V^{-1} \thinspace s^{-1}$, respectively, which are similar to the transport results \cite{PhysRevB.50.5180, PhysRevB.32.1183}.
The high $\mu$'s support the suppressed back-scattering of the topological bands of $\rm{SrAs_3}$.

Next, we investigate the interband conductivity $\sigma_{1}^{\mathrm{IB}}(\omega)$.
For $E \parallel k_x$, $\sigma_{1}(\omega)$ displays an absorption peak (A) at 30 meV and a flat conductivity (B) over the 70 - 129 meV range.
At higher energies, $\sigma_{1}(\omega)$ gradually increases (C) with energy.
For $E \parallel k_z$, 
the flat optical conductivity $\sigma_{zz}^{\mathrm{flat}} \approx 239 \ \mathrm{cm^{-1} \thinspace \mathrm{\Omega^{-1}}}$ is significantly enhanced over $\sigma_{xx}^{\mathrm{flat}} \approx 83 \ \mathrm{cm^{-1} \thinspace \mathrm{\Omega^{-1}}}$ for $E \parallel k_x$. 
In the region C, $\sigma_{1}(\omega)$ drops markedly for $E \parallel k_z$ around $\hbar\omega = 129$ meV in contrast to the increase for $E \parallel k_x$.
To understand the behaviors of A, B, and C both qualitatively and quantitatively, we perform a theoretical calculation using a four-band model Hamiltonian that takes into account the two crossing bands of the nodal ring including their spin degrees of freedom:
\begin{align} \label{eq:ham_full}
    H(\bm{k}) = f_0(\bm{k})\sigma_0 s_0 + f_1(\bm{k})\sigma_1 s_0
    + f_2(\bm{k})\sigma_2 s_0 + \Delta_{\rm SOC}\sigma_3 s_3,
\end{align}
where $f_0(\bm{k}) = a_0 + a_x k_x^2 + a_{xy} k_x k_y + a_y k_y^2 + a_z k_z^2$, $f_1(\bm{k}) = b_0 + b_x k_x^2 + b_{xy} k_x k_y + b_y k_y^2 + b_z k_z^2$, $f_2(\bm{k}) = \hbar v_z k_z$, and $\bm{k}$ is the wave vector measured from the Y point in the BZ, and $\bm{\sigma}$ and $\bm{s}$ are Pauli matrices acting on the pseudospin and spin degrees of freedom, respectively.
$\Delta_{\rm SOC}$ is the strength of the SOC, which, as a leading approximation, we take as constant. Here the coefficients $a_0$, $a_i$, $b_i$ ($i=x$, $xy$, $y$, $z$), and $v_z$ are obtained from first-principles calculations. For the band overlap energy 2$\varepsilon_0 = 2b_0$ and SOC $\Delta_{\mathrm{SOC}}$, which are difficult to estimate correctly from first-principles calculations, we determine them by comparing with the optical data (see SM Sec. III \cite{SM}). Then we calculate the interband conductivity $\sigma_1(\omega)$ using the Kubo formula. The full numerical results of $\sigma_1(\omega)$ are presented in Fig. 3(b).

To gain some insight into the optical features, we adopt a simple picture that the nodal ring is considered as a collection of gapped anisotropic graphene sheets, as shown in Fig. 3(d). In this picture, the Hamiltonian in Eq. (1) at low frequencies can be transformed into \cite{SM}
\begin{align} \label{eq:gr_ham}
    H(\bm{k}) = \Delta_{\mathrm{tilt}}(\bm{k})\sigma_0 + \hbar v_{\rho}(\phi) \delta k_{\rho} \sigma_1 + \hbar v_z k_z \sigma_2 + \Delta_{\mathrm{SOC}} \sigma_3. \nonumber \\
\end{align}
Here $(k_{\rho}, \phi)$ are the polar coordinates of $(k_x, k_y)$, $\delta k_{\rho}$ and $\delta k_z$ are wave vectors measured from the nodal ring along the radial and $k_z$ directions, respectively, and $v_{\rho}(\phi)$ and $v_z$ are the corresponding Fermi velocities along the two directions. Note that $v_{\rho}(\phi)$ has an angular dependence, while $v_z$ is constant. $\Delta_{\mathrm{tilt}}(\bm k)$ is the energy tilt term and $\Delta_{\mathrm{SOC}}$ acts as a mass term opening an energy gap.

Then the optical conductivity of the nodal ring can be obtained by summing up all the contributions from each gapped anisotropic graphene sheet along the nodal ring as
\begin{equation} \label{eq:opcd_as_sum}
    \sigma_{ii} (\omega) = k_0 \int_0^{2\pi} \frac{d\phi}{2\pi} \ \sigma_{1}^{\mathrm{ gr}} (\omega, \phi) \mathcal{F}_{ii} (\phi),
\end{equation}
where $i = x$, $y$, $z$ are spatial directions, $\sigma_{ii}^{\mathrm{gr}}(\omega, \phi)$ is the optical conductivity of the gapped anisotropic graphene, $k_0$ is the characteristic scale of the nodal ring size which will be defined later, and $\mathcal{F}_{ii}(\phi)$ are geometric factors related to the shape of the nodal ring (see SM Sec. IV \cite{SM}).

The optical conductivity for the Hamiltonian in Eq. (\ref{eq:gr_ham}) at zero temperature in the clean limit is then given by \cite{SM}
\begin{align} \label{eq:opcd}
    \sigma_{ii}(\omega) &\approx \sigma^{\mathrm{flat}}_{ii} \left[1+\left(\frac{2\Delta_{\mathrm{SOC}}}{\hbar\omega}\right)^2\right] \Theta(\hbar\omega - 2 \Delta_{\mathrm{SOC}}).
\end{align}
Here, $\sigma_{ii}^{\mathrm{flat}}$ is the flat optical conductivity which will be discussed in Eq. (\ref{eq:opcd_flat_analytic}), the term inside the square bracket represents the effect of SOC, and $\Theta(x)$ is the Heaviside step function. The result in Eq. (\ref{eq:opcd}) shows that, at the onset of interband transitions, the SOC-induced optical peak appears at $\hbar\omega = 2 \Delta_{\mathrm{SOC}}$ in both $\sigma_{xx}$ and $\sigma_{zz}$, and therefore the peak position defines the magnitude of the SOC gap.

In obtaining Eq. (\ref{eq:opcd}), we use the fact that $E_{\mathrm{F}}$ of $\mathrm{SrAs_3}$ lies within the SOC gap for at least a certain range of angle, as shown in Fig. 3(c), giving a peak corresponding to the gap size. Note that Eq. (\ref{eq:opcd}) is an approximate form neglecting the effect of the energy tilt term. For details, see SM Sec. IV \cite{SM}.

Based on the experimental data in Fig. 3(a), we estimate 2$\Delta_{\mathrm{SOC}} \cong 30$ meV. The SOC gap is one of the key factors that determines the band structure of NLSMs but has seldom been measured precisely due to its small magnitude. We emphasize that the high energy resolution of IR spectroscopy, in combination with the successful theoretical analysis, allows for the reliable determination of $\Delta_{\mathrm{SOC}}$.

As frequency increases, the optical conductivity approaches $\sigma^{\mathrm{flat}}_{ii}$ in Eq. (\ref{eq:opcd}). This can be obtained analytically as \cite{SM}
\begin{subequations} \label{eq:opcd_flat_analytic}
    \begin{align} 
        \sigma_{xx}^{\mathrm{flat}} &= k_0 \frac{e^2}{16\hbar} \frac{g}{4} \frac{v}{v_z} \left[\frac{k_s}{k_l}+\frac{k_l}{k_s} + \left(\frac{k_s}{k_l}-\frac{k_l}{k_s}\right) \cos{\left(2\phi_0\right)} \right], \\
        \sigma_{zz}^{\mathrm{flat}} &= k_0 \frac{e^2}{16\hbar} g \frac{v_z}{v},
    \end{align}
\end{subequations}
where $k_l$ and $k_s$ are the lengths of the semi-major and semi-minor axes of the nodal ring, $k_0 = \sqrt{k_l k_s}$, $v = \hbar k_0 / m_0$, $\phi_0$ is the angle between the semi-major axis of the nodal ring and the $k_x$ axis, $g=2$ is the band degeneracy factor. From $k_0=0.068 \ \mathrm{\AA^{-1}}$, $k_l/k_s=1.16$, $v=2.88 \times 10^5 \ \mathrm{m/s}$, $v_z=3.48 \times 10^5 \ \mathrm{m/s}$, and $\phi_0 = 23.5^{\circ}$ \cite{SM}, $\sigma_{xx}^{\mathrm{flat}}$ and $\sigma_{zz}^{\mathrm{flat}}$ are estimated to be 78 $\mathrm{cm}^{-1}\mathrm{\Omega}^{-1}$ and 247 $\mathrm{cm}^{-1}\mathrm{\Omega}^{-1}$, respectively, showing good agreement with the experimental result.

It should be noted that for a given nodal ring size ($k_0 = \sqrt{k_l k_s}$), $\sigma_{xx}^{\mathrm{flat}}$ depends on how much the nodal ring is elongated and rotated, taking into account the mixing between the semi-major and semi-minor directions, while $\sigma_{zz}^{\mathrm{flat}}$ does not depend on the nodal ring shape and the rotation angle. For a circular nodal ring ($k_l=k_s$) with equal velocity $v = v_z$, Eq. (\ref{eq:opcd_flat_analytic}) reduces to $\sigma_{xx}^{\mathrm{flat}} = \sigma_{zz}^{\mathrm{flat}}/2 =  ge^2 k_0/(32\hbar)$ consistent with the previous result \cite{PhysRevLett.119.147402}.

Having investigated the SOC-induced peak (A) and the flat conductivity (B), we move to the region C. For $\hbar \omega$ $>$ $2\varepsilon_0 \approx$ 129 meV, the flat conductivity is no longer maintained and $\sigma_{xx}$ increases gradually while $\sigma_{zz}$ shows a sudden drop, indicating that the picture we employed in Eq. (2) does not hold any more at high frequencies \cite{PhysRevLett.119.147402, PhysRevB.96.155150}. At $\hbar \omega = 2\varepsilon_0$, the joint density of states for interband transitions shows an abrupt change. Above this frequency, the optical conductivity exhibits a non-trivial frequency-dependence depending on the direction due to the difference in the velocity matrix elements along the $k_x$ and $k_z$ directions.
Remarkably, our analytic estimation for the flat conductivity using Eq. (\ref{eq:opcd_flat_analytic}) and the full numerical calculation using the Kubo formula over a wide frequency range are in excellent agreement with the experiment not only qualitatively but also quantitatively.

\begin{figure}[!b]
\includegraphics[width=1.0\columnwidth]{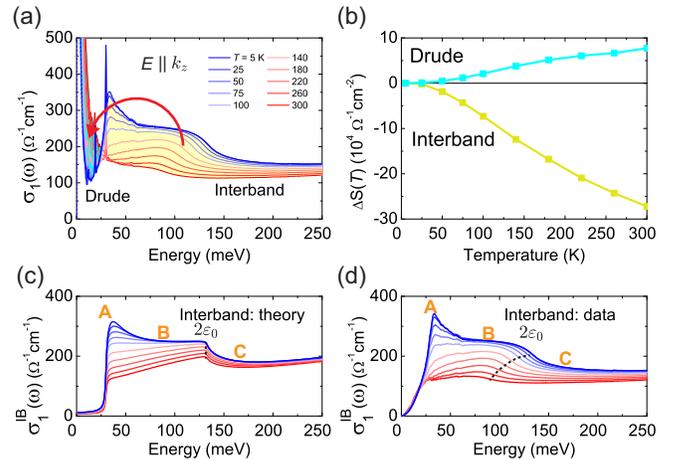} 
\caption{\label{fig:fig4}
(a) $T$-dependent $\sigma_{1}(\omega)$ for 
 5 K $\leq T \leq$ 300 K. The two shaded areas visually show that the optical spectral weights change with $T$. (b) Optical spectral weight difference $\Delta S (T) = S (T) - S (5\,\,\mathrm{K})$ for the Drude and 
IB transition. 
(c) and (d) show the numerical and experimental IB conductivity $\sigma_{1}^{\mathrm{IB}}(\omega)$, respectively. In (d), the dashed curve highlights the shift of 2$\varepsilon_0$.
}
\end{figure}

We now move on to the $T$-dependent characteristics of $\sigma_{1}(\omega)$. Figure 4(a) illustrates that with increasing temperature, the Drude peak becomes broader and gains spectral weight while the interband transition decreases and loses spectral weight.
This redistribution of spectral weight can be understood through two factors. Firstly, the thermal population of electrons in the conduction band (CB) and holes in the valence bands (VB) leads to a broadening of the Fermi-Dirac distribution. Secondly, the shift in chemical potential resulting from density conservation reduces the regions where interband transitions are allowed (See SM Sec. IV-2, V \cite{SM}). These two factors collectively contribute to the 
enhancement of the Drude weight and the suppression of interband transitions.
For quantitative analysis, we calculate the optical spectral weight of the Drude peak, $S^{\mathrm{D}} = \int_{0}^{\infty} \sigma_{1}^{\mathrm{D}}(\omega)d\omega = \frac{\omega_p^2}{8} $ and interband transition $S^{\mathrm{IB}} = \int_{0}^{\omega_c} \sigma_{1}^{\mathrm{IB}}(\omega)d\omega$.
In calculating  $S^{\mathrm{IB}}$, the cutoff frequency $\omega_c$  was taken to be 0.7 eV (see SM Sec II-3 \cite{SM}). 
In Fig. 4(b), we display the $T$-driven changes of the spectral weight  $\Delta S (T) = S (T) - S (5 \,\,\mathrm{K})$ for the Drude peak and IB transition.
We find $\Delta S^{\mathrm{D}} >$ 0 for the Drude and $\Delta S^{\mathrm{IB}} <$ 0 for the IB, indicating that the optical spectral weight is transferred from the IB to the Drude with increasing temperature.
One interesting observation is that $|\Delta S^{\mathrm{IB}}|$ is substantially larger than $\Delta S^{\mathrm{D}}$, as seen by comparing the two shaded spectral areas in Fig. 4(a).
The non-conservation of the total spectral weight, $\Delta S^{\mathrm{D}}$ + $\Delta S^{\mathrm{IB}}$ $\neq 0$, seemingly violates  the optical sum rule.
Similar behavior was reported in other materials such as FeSi, YBa$_2$Cu$_3$O$_x$, La$_{2-x}$Sr$_x$CuO$_4$ \cite{PhysRevLett.71.1748, PhysRevB.63.134514, PhysRevLett.91.037004}.
The origin of their sum rule violation is poorly understood. The lost spectral weight is expected to be recovered in a frequency range not probed experimentally \cite{PhysRevLett.71.1748}.
This notion indicates the necessity of expanding the optical measurement on $\rm{SrAs_3}$ beyond the energy ranges investigated in the present study.

For further analysis, we numerically  calculated the $T$-dependent $\sigma_{1}^{\mathrm{IB}}(\omega)$ by incorporating the thermal effect, i.e., the Fermi-Dirac distribution and $T$-driven chemical potential $\mu(T)$ shift in the Kubo formula (see SM Sec. V \cite{SM}). 
Both the theoretical and experimental results show consistently that $\sigma_{1}^{\mathrm{IB}}(\omega)$ decreases and $\sigma_{1}^{\mathrm{flat}}$ becomes less distinct with increasing $T$ as shown in Figs. 4(c) and 4(d). However, there is a notable difference between the two: in the experimental data, $\sigma_{1}^{\mathrm{IB}}(\omega)$ shows a redshift of 2$\varepsilon_0$, while in the numerical results, 2$\varepsilon_0$ remains constant.
This discrepancy suggests the presence of an additional factor beyond thermal population effects, possibly indicating a $T$-driven evolution of the band structure not captured in the numerical calculations.
However, at this stage, our understanding of this phenomenon is still limited.
Further study is required to elucidate the exact nature of the $T$-dependent $\sigma_{1}(\omega)$.  

To conclude, we performed the measurement of optical transitions on a nodal ring semimetal by adopting $\rm{SrAs_3}$, a unique material that possesses only a single nodal ring in the BZ without any interrupting trivial bands. 
By combining model and first-principles calculations, our axis-resolved optical measurements revealed fundamental properties of NLSMs beyond the flat absorption, such as the SOC gap, the band overlap energy, and the intrinsic geometric features of the nodal ring.
Our study demonstrates that $\sigma_{1}(\omega)$ originates exclusively from the single nodal ring, establishing $\rm{SrAs_3}$ as an ideal platform for investigating nodal rings in topological materials.

\begin{acknowledgments}
This work was supported by the NRF grant funded by the Korea government (NRF-2021R1A2C1009073).
J.J. was supported by Basic Science Research Program of Korea (NRF-2018R1A6A1A06024977).
J.J. and H.M. were supported by the NRF grant funded by the Korea government (NRF-2018R1A2B6007837 and NRF-2023R1A2C1005996), the Creative-Pioneering Researchers Program through Seoul National University (SNU), and the Center for Theoretical Physics.
H. K. and J. S. K. were supported by the Institute for Basic Science (IBS) through the Center for Artificial Low Dimensional Electronic Systems (no. IBS-R014-D1), and by the NRF through the Basic Science Research Program (2022R1A2C3009731), and the Max Planck POSTECH/Korea Research Initiative (2022M3H4A1A04074153).
T.P. and J.S. were supported by the NRF grant funded by the Korea government (NRF-2021R1A2C2010972).
The work at HYU was supported by the NRF grant funded by the Korean government (MSIT) (2022R1F1A1072865 and RS-2022-00143178) and BrainLink program funded by the Ministry of Science and ICT through the NRF (2022H1D3A3A01077468). 
\end{acknowledgments}

\bibliographystyle{apsrev4-2}
\bibliography{main}
\end{document}


\title{Supplemental Material for ``Optical transitions of a single nodal ring in SrAs$_3$: radially and axially resolved characterization''}

\author{Jiwon Jeon}
 \thanks{These authors contributed equally to this work.}
\affiliation{%
 Natural Science Research Institute, University of Seoul, Seoul 02504, Korea
 }%
 \affiliation{%
 Physics Department, University of Seoul, Seoul 02504, Korea
 }%

\author{Jiho Jang}%
 \thanks{These authors contributed equally to this work.}
 \affiliation{%
 Department of Physics and Astronomy, Seoul National University, Seoul 08826, Korea
 }%

\author{Hoil Kim}%
 \affiliation{%
 Center for Artificial Low Dimensional Electronic Systems, Institute for Basic Science (IBS), Pohang 37673, Korea
 }%
  \affiliation{%
 Department of Physics, Pohang University of Science and Technology (POSTECH), Pohang 37673, Korea
 }%

\author{Taesu Park}%
 \affiliation{%
 Department of Chemistry, Pohang University of Science and Technology (POSTECH), Pohang 37673, Korea
 }%

 \author{DongWook Kim}%
 \affiliation{%
 Department of Physics, Hanyang University, Seoul 04763, Korea
 }%

 \author{Soonjae Moon}%
 \affiliation{%
 Department of Physics, Hanyang University, Seoul 04763, Korea
 }%

\author{Jun Sung Kim}%
 \affiliation{%
 Center for Artificial Low Dimensional Electronic Systems, Institute for Basic Science (IBS), Pohang 37673, Korea
 }%
  \affiliation{%
 Department of Physics, Pohang University of Science and Technology (POSTECH), Pohang 37673, Korea
 }%

\author{Ji Hoon Shim}%
 \affiliation{%
 Department of Chemistry, Pohang University of Science and Technology (POSTECH), Pohang 37673, Korea
 }%

\author{Hongki Min}%
 \email{hmin@snu.ac.kr}
 \affiliation{%
 Department of Physics and Astronomy, Seoul National University, Seoul 08826, Korea
 }%

\author{Eunjip Choi}%
\email{echoi@uos.ac.kr}
 \affiliation{%
 Physics Department, University of Seoul, Seoul 02504, Korea
 }%

\date{\today}
\newcommand{\beginsupplement}{%
        \setcounter{table}{0}
        \renewcommand{\thetable}{S\arabic{table}}%
        \setcounter{figure}{0}
        \renewcommand{\thefigure}{S\arabic{figure}}%
     }



\maketitle
\beginsupplement
\tableofcontents

\newpage
\section {\uppercase\expandafter{\romannumeral1}. Sample preparation and experimental methods}

Single crystals of $\rm{SrAs_3}$ with a typical size of 2 mm $\times$ 3 mm $\times$ 0.5 mm were grown using the Bridgman method, and their structure was characterized through X-ray Diffraction (XRD) and 
Scanning Tunneling Electron Microscopy (STEM) \cite{Kim2022}.   
The transport properties of $\rm{SrAs_3}$ were characterized through 
dc-resistivity, Hall-resistivity, and Shubnikov-de Haas (SdH) oscillation measurements. The Fermi level and the carrier density of the electron and hole,  which change from sample to sample, were determined from the transport results
for each crystal.  Angle-dependent magneto-resistivity measurements were performed at low-temperature and high-magnetic field to investigate the characteristic quantum  transport properties associated with the single nodal ring, and were interpreted in terms of the Berry flux and weak antilocalization. The electronic band structure was probed by an Angle-Resolved Photoemission Spectroscopy (ARPES) measurement.

The optical reflectivity of $\rm{SrAs_3}$ was measured in the infrared range (from 6.2 meV to 0.8 eV) for various temperatures ranging from 5 K to 300 K. For the axis-resolved reflectivity measurements, three types of linear polarizers working in the (1) Far-infrared range (0.12 meV $\sim$ 62 meV), (2) Mid-infrared range (35 meV $\sim$ 620 meV), and (3) Near-infrared range (120 meV $\sim$ 1.2 eV) were employed, respectively. The three sets of data were combined into one before applying the Kramers-Kronig constrained fitting.
This reflectivity measurement was performed using a Fourier-transform infrared spectrometer (Vertex 70V) in combination with the in-situ gold evaporation technique \cite{Homes:93}.
The high-energy optical reflectivity was measured at $T$ = 300 K in the range from 0.7 eV to 4.5 eV for the axial and radial directions of the nodal ring, respectively, utilizing the polarized light of a Spectroscopic Ellipsometer (J.A. Woollam VASE).
\newpage

\section {\uppercase\expandafter{\romannumeral2}. Temperature-dependent reflectivity and optical conductivity}
\subsection{\uppercase\expandafter{\romannumeral2}-1. Optical reflectivity data}

\begin{figure}[h]
\caption{\label{fig:sm:reflectivity} Temperature dependent reflectivity for $E \parallel k_x$ and $E \parallel k_z$.
}
\includegraphics[width=1\columnwidth]{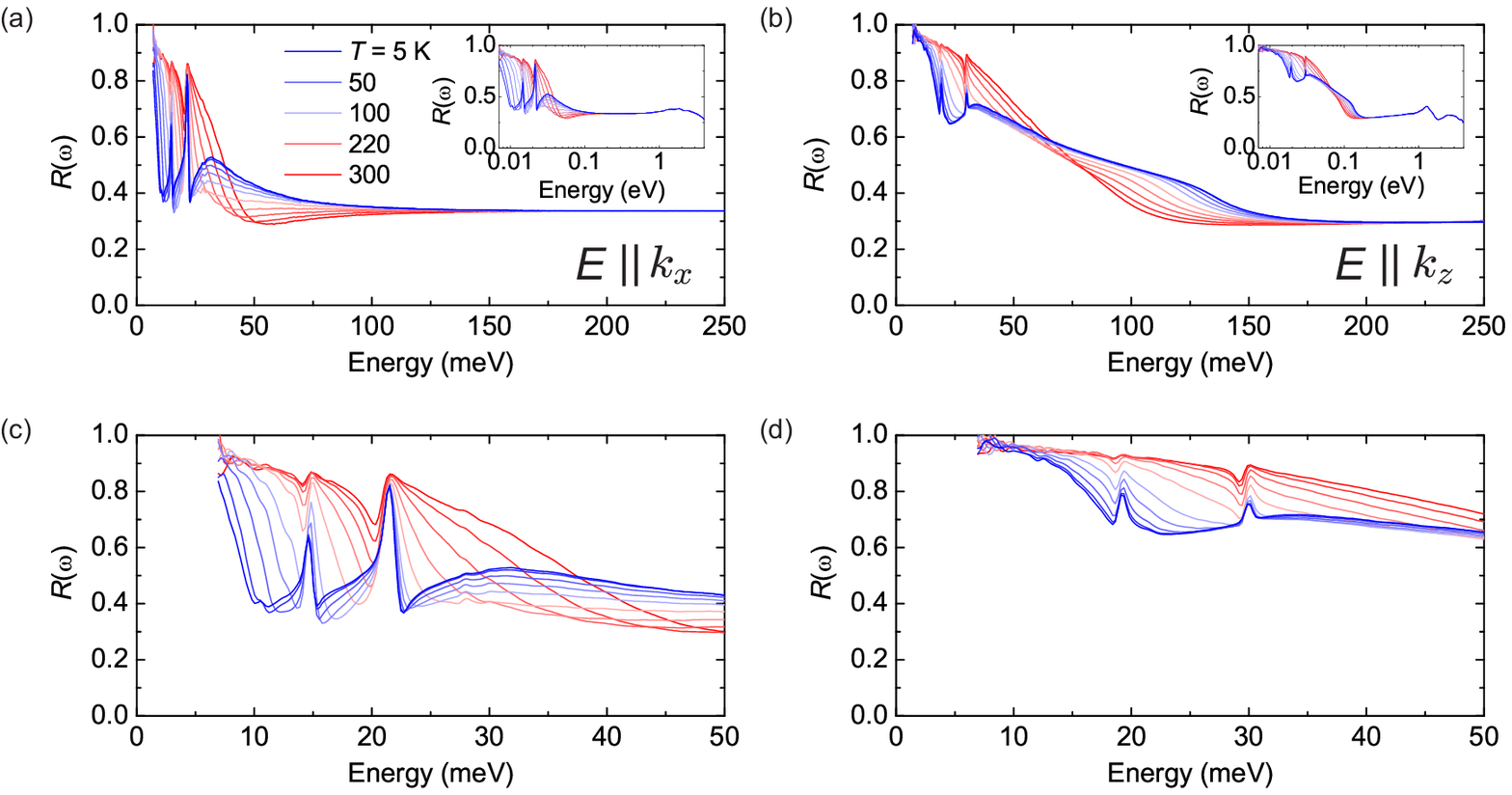}
\end{figure}

The optical reflectivity was measured with incident light that is polarized along $E$ $\parallel$ $a$ and $E$ $\parallel$ $b$.
In the $k$-space, they correspond to $E$ $\parallel$ $k_x$ and $E$ $\parallel$ $k_z$, which will probe the radial and axial directions of the nodal ring, respectively. 
Figure \ref{fig:sm:reflectivity} shows that, for the $E$ $\parallel$ $k_x$ polarization, reflectivity rises with decreasing energy for $\hbar\omega$ $<$ 50 meV ($T$ = 300 K) due to a Drude response. 
As $T$ decreases, the plasma edge of the Drude peak shifts to a lower energy.
In addition, a new hump develops at 30 meV for $T$ $<$ 140 K.
For $E$ $\parallel$ $k_z$, the Drude reflectivity is broader and its level is higher than $E$ $\parallel$ $k_x$.
The 30 meV-hump appears in $E$ $\parallel$ $k_z$ as well.   
Notably, a shoulder-like feature is present at $\sim$ 90 meV which is absent for $E$ $\parallel$ $k_x$.
As $T$ decreases, the shoulder feature shifts to higher energy reaching $\sim$ 140 meV at $T$ = 5 K. 
Insets show the reflectivity measured over wide energy at $T$ = 300 K.

\newpage
\subsection{\uppercase\expandafter{\romannumeral2}-2. Fit of $\sigma_1(\omega)$ using the Drude, phonon, and interband contributions}

\begin{figure}[h]
\caption{\label{fig:sm:fitcomponent} Fitting of the Drude, phonon, and interband components using $\sigma_1(\omega)$.}
\includegraphics[width=1\columnwidth]{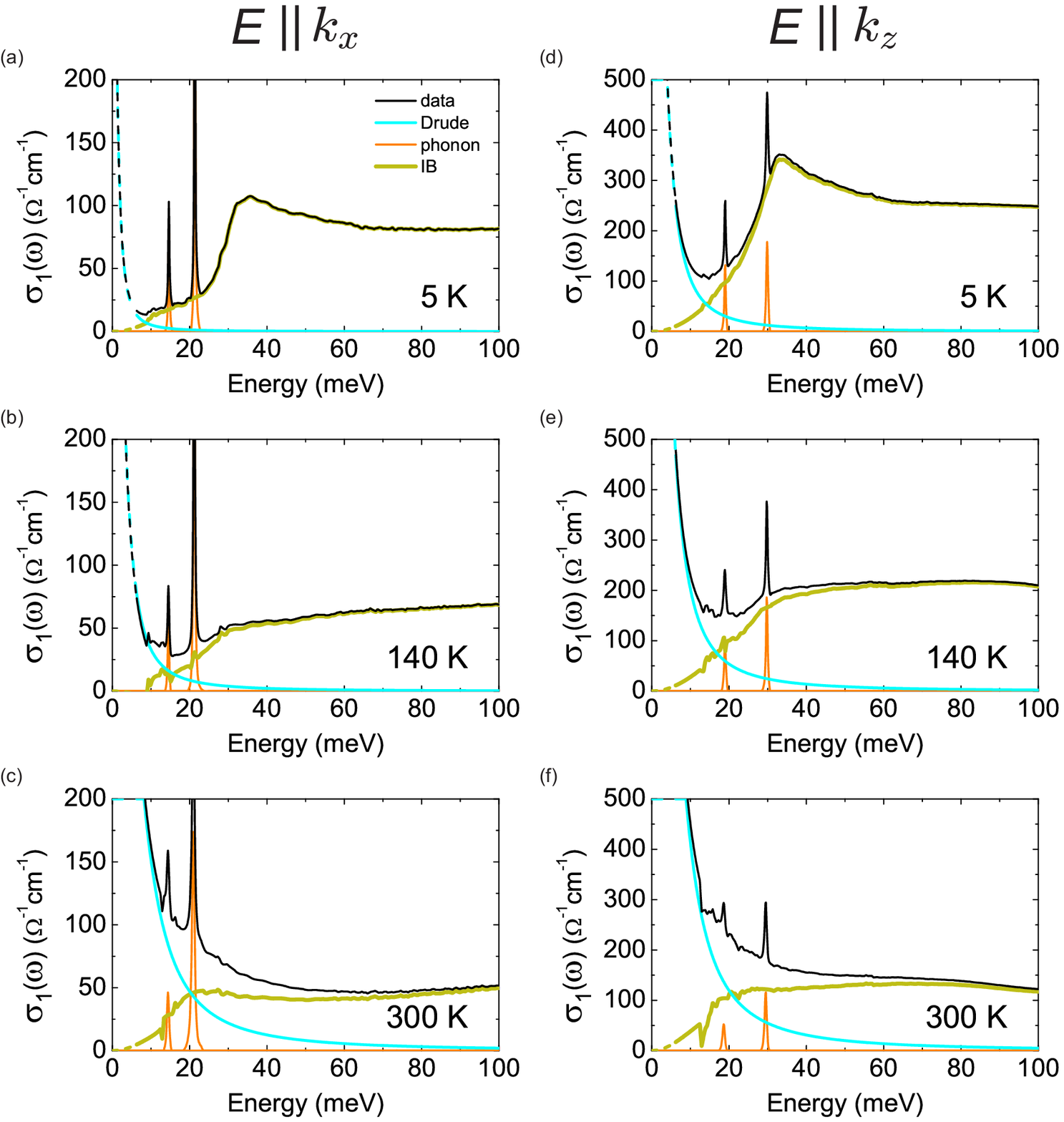}
\end{figure}

Figure \ref{fig:sm:fitcomponent} shows that $\sigma_1(\omega)$ consists of  Drude, phonon, and interband contributions as $\sigma_1(\omega) = \sigma_{1}^{\mathrm{D}}(\omega) + \sigma_{1}^{\mathrm{Ph}}(\omega) + \sigma_{1}^{\mathrm{IB}}(\omega)$, where $\sigma_{1}^{\mathrm{D}}(\omega) = \frac{\omega^2_{p}}{4\pi}\frac{\gamma}{(\gamma^2+\omega^2)}$ is the Drude conductivity, $\sigma_{1}^{\mathrm{Ph}}(\omega) = \displaystyle\sum_{j=1}^{2}\frac{\omega^2_{p,j}\omega^2\gamma_j}{(\omega^2_{0,j}-\omega^2)^2+\omega^2\gamma_{j}^2}$ is the phonon contribution described by Lorentzian peaks.
Here, the fitting parameters $\omega_{p}$ and $\gamma$ are the plasma frequency and width of the Drude peak. $\omega_{p,j}$, $\omega_{0,j}$ and $\gamma_{j}$ are the plasma frequencies, positions, and widths of the two phonon peaks, respectively, for $j=1$ and $2$.
For the interband conductivity, $\sigma_{1}^{\mathrm{IB}}(\omega)$ was obtained by subtracting $\sigma_{1}^{\mathrm{D}}(\omega)$ and $\sigma_{1}^{\mathrm{Ph}}(\omega)$ from $\sigma_{1}(\omega)$, $\sigma_{1}^{\mathrm{IB}}(\omega)$ = $\sigma_{1}$$(\omega)$ $-$ [$\sigma_{1}^{\mathrm{Ph}}(\omega)$ $+$ $\sigma_{1}^{\mathrm{D}}(\omega)$] 
\newline

\begin{figure}[h]
\caption{\label{fig:sm:seperate} Fitting curves for the Drude, phonon, and interband conductivities}
\includegraphics[width=1\columnwidth]{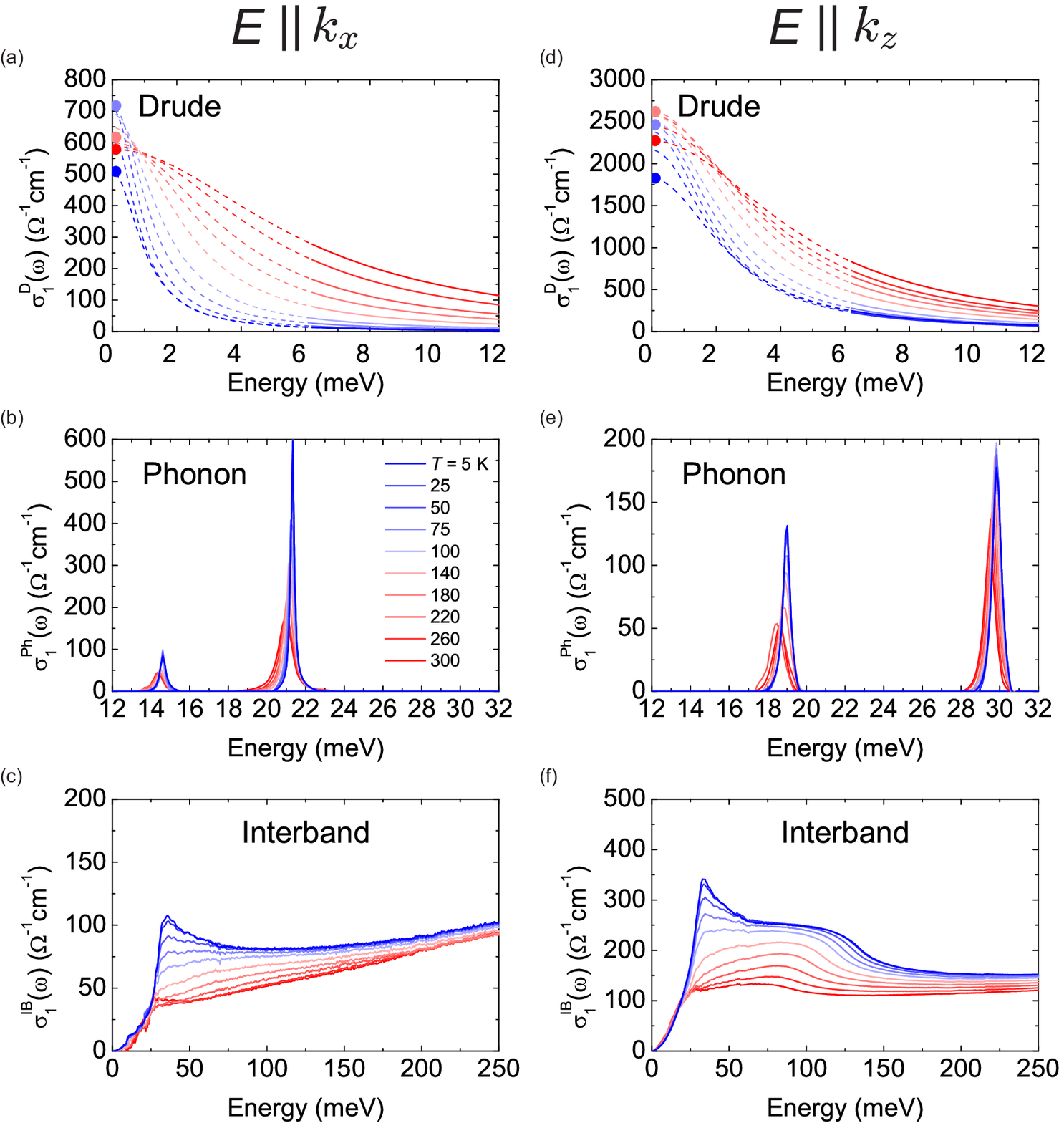}
\end{figure}

\begin{figure}[h]
\caption{\label{fig:sm:drude} Drude parameters for 5 K $\leq$ $T$ $\leq$ 300 K.}
\includegraphics[width=1\columnwidth]{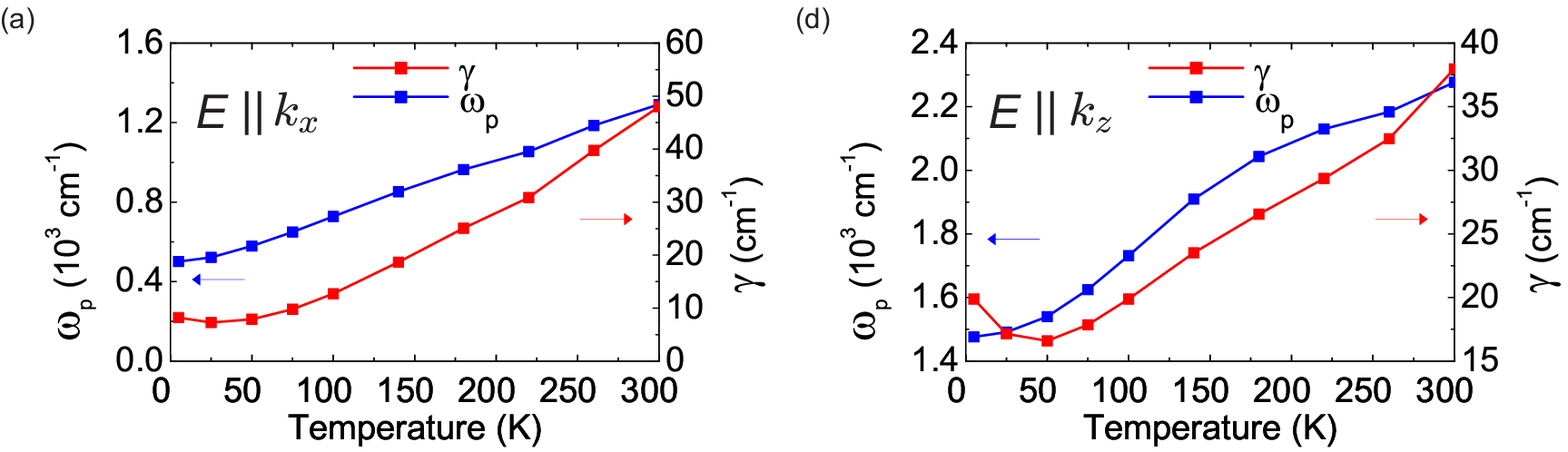}
\end{figure}

Figure \ref{fig:sm:seperate} displays the fitting curves $\sigma_{1}^{\mathrm{D}}(\omega)$, $\sigma_{1}^{\mathrm{Ph}}(\omega)$ and $\sigma_{1}^{\mathrm{IB}}(\omega)$ obtained from the fit.
In Figs. \ref{fig:sm:seperate}(a) and \ref{fig:sm:seperate}(d), the optical conductivity below 6.2 meV is an extrapolation derived from the constrained Kramers-Kronig analysis (dashed lines), showing a good agreement with the DC-conductivity obtained from transport measurement (filled symbols) at $\omega = 0$.
Figure \ref{fig:sm:drude} shows the temperature dependence of the fitting parameters $\omega_{p}^{2}$ and $\gamma$ for the Drude peak.
Table \ref{table:stable1} summarizes the Drude and phonon fitting parameters at $T$ = 5 K for the two light polarizations. 
\newline

\newpage

\newpage

\newpage

\begin{table}[h]
\centering
\caption{\label{table:stable1}Drude and Phonon fitting parameters for $T$ = 5 K}
\setlength{\tabcolsep}{10pt}
\label{t4}
\begin{tabular}{c|cc|ccc|ccc}
\noalign{\smallskip}\noalign{\smallskip}\hline\hline
\multirow{2}{*}{(cm$^{-1}$)} & \multicolumn{2}{c|}{Drude} & \multicolumn{3}{c|}{Phonon 1} & \multicolumn{3}{c}{Phonon 2}\\
\cline{2-9}
      & $\omega_{p}$  & $\gamma$ & $\omega_0$ & $\omega_p$ & $\gamma$ & $\omega_0$ & $\omega_p$ & $\gamma$\\
\hline
 $E \parallel k_x$ & 499.9 & 8.2 & 117.8 & 126.4 & 3.1 & 171.9 & 242.7 & 1.7 \\
 $E \parallel k_z$ & 1476.2 & 19.9 & 153.1 & 163.0 & 3.3 & 240.8 & 228.0 & 4.6 \\
\hline
\hline
\end{tabular}
\end{table}

\newpage
\subsection{\uppercase\expandafter{\romannumeral2-3:}  Optical spectral weight transfer}

\begin{figure}[h]
\caption{\label{fig:sm:sum_rule} Spectral weight transfer and optical sum rule
}
\includegraphics[width=1\columnwidth]{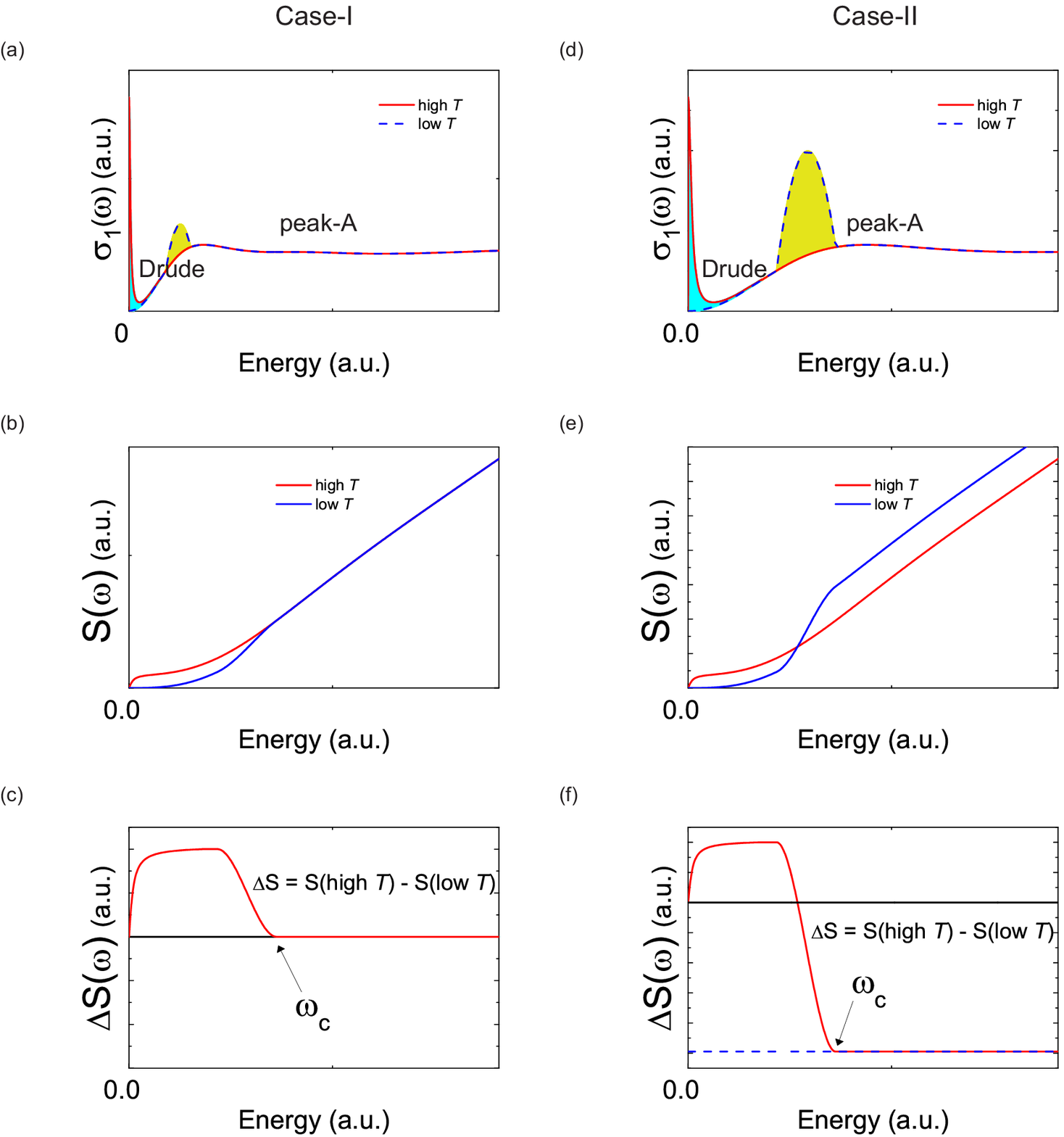}
\end{figure}

While the optical sum rule holds against $T$-variation in numerous materials, it is seemingly violated in some materials.
In Fig. \ref{fig:sm:sum_rule}, we schematically illustrate two idealized cases in which the optical sum rule is satisfied (Case-I) and not satisfied (Case-II), respectively.
Here, an interband peak (A) is suppressed and Drude peak (D) grows with increasing $T$.
These $T$-dependent changes occur at $\omega < \omega_c$ and are exhausted at $\omega_c$. 
In Case-I, the optical spectral weight lost in peak-A is exactly recovered by the weight gained in D, conserving the total optical spectral weight. 
The spectral weight functions $S(T, \omega) = \int_{0}^{\omega} \sigma_{1}(\omega')d\omega'$ for the low-$T$ and high-$T$ merge at a certain frequency $\omega_{c}$. The spectral weight difference $\Delta S (\omega)$ = $S(\omega, \mathrm{high}\,\,T ) - S(\omega, \mathrm{low}\,\,T )$ vanishes at $\omega_{c}$ which is taken as the cutoff frequency for calculating the corresponding interband spectral weight.

Case-II illustrates the case in which, unlike Case-I, the spectral weight lost in peak-A is larger than that gained in D. The total spectral weight is not conserved within the relevant energy range.
Here the two $S(T, \omega)$ curves do not merge at $\omega_c$. Instead, $\Delta S (\omega)$ becomes  non-zero and maintains this constant value above the cutoff energy $\omega_{c}$.

\begin{figure}[h]
\caption{\label{fig:sm:deltaS} Spectral weight difference $\Delta S (\omega)$ for (a) $E \parallel k_x$ and (b) $E \parallel k_z$.
}
\includegraphics[width=1\columnwidth]{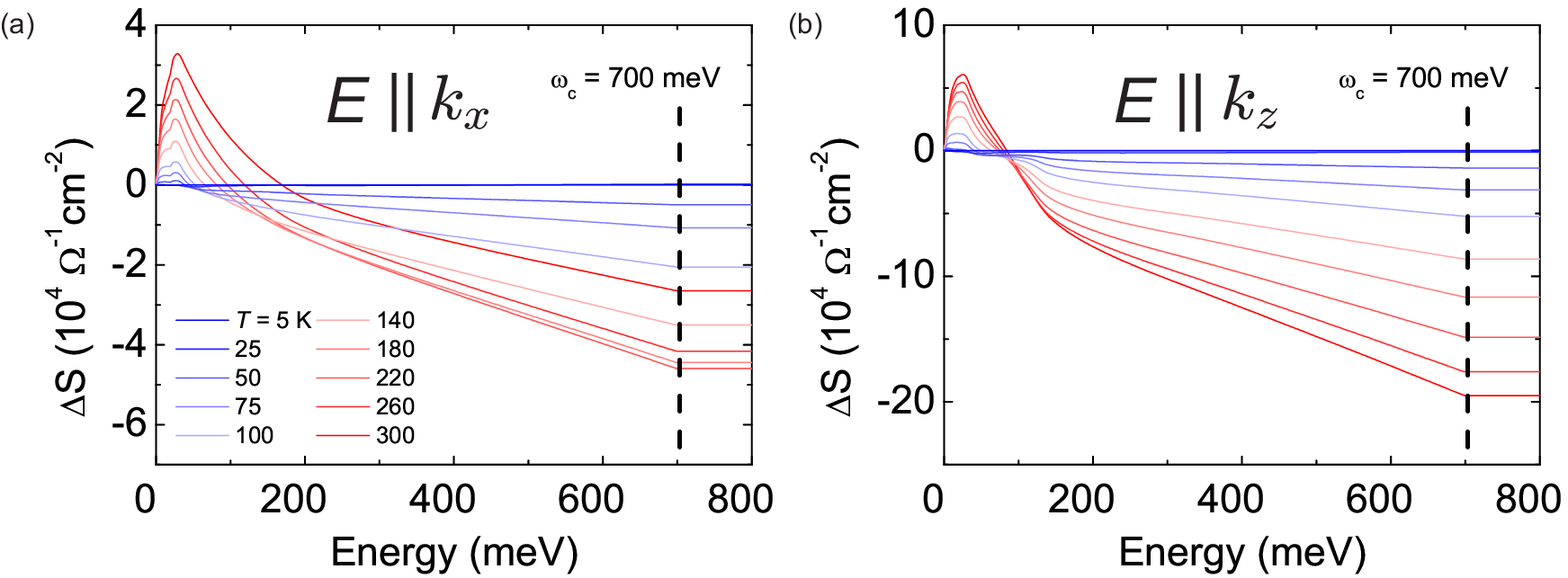}
\end{figure}

On the basis of this analysis, we examine $\Delta S (\omega)$ of $\mathrm{SrAs_3}$ for $E \parallel k_x$ and $E \parallel k_z$. 
In Fig. \ref{fig:sm:deltaS},  $\Delta S (\omega)$ shows similar behavior to Case-II, demonstrating the non-conservation of the optical spectral weight. The cutoff frequency is taken as $\omega_c$ = 700 meV.

\newpage
\begin{figure}[h]
\caption{\label{fig:sm:fig4_xx} Optical spectral weight analysis for $E \parallel k_x$.
}
\includegraphics[width=1\columnwidth]{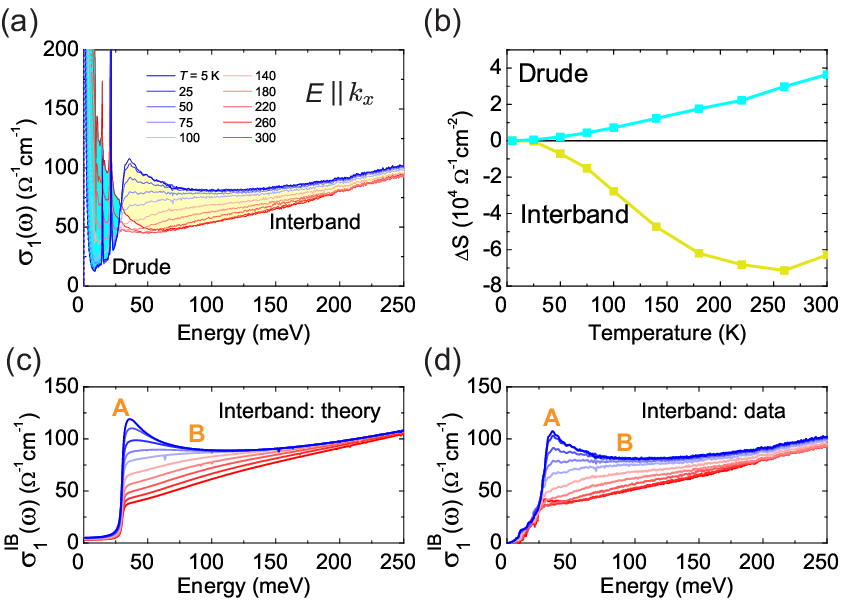}
\end{figure}

(a) Optical conductivity for $E \parallel k_x$ over 5 K $\leq T \leq$ 300 K.
The Drude (interband) contribution to $\sigma_{1}(\omega)$ increases (decreases) with increasing temperature. The two shaded areas illustrate their optical spectral weight changes.
(b) $T$-dependent spectral weight change $\Delta S (T) = S (T) - S (5\,\,\mathrm{K})$ for the Drude and interband transition. 
(c) and (d) show the numerical and experimental interband conductivities $\sigma_{1}^{\mathrm{IB}}(\omega)$, respectively. 

For $E \parallel k_x$, the Drude and interband conductivity increases and decreases similarly to $E \parallel k_z$.
The optical spectral weight is transferred from the interband to Drue with $T$. The $T$-dependent spectral weights displayed in (b) show similar behavior as in $E \parallel k_z$.

\newpage
\section{\uppercase\expandafter{\romannumeral3}. Model and calculation methods}
\subsection{\uppercase\expandafter{\romannumeral3}-1. Band structures and first-principles calculations details} \label{sec:dft}
The electronic structure of the bulk $\mathrm{SrAs_3}$ is calculated using the density functional theory (DFT) calculation implemented in WIEN2K \cite{doi:10.1063/1.5143061}, using a full-potential linearized augmented plane wave method. The exchange-correlation functional of the modified Becke-Johnson (mBJ) potential is used for an accurate description of the valence and conduction bands in semimetallic $\mathrm{SrAs_3}$, as well as to avoid the bandgap underestimation of the Perdew-Berke-Ernzerhof (PBE) functional \cite{PhysRevLett.102.226401}. The core separation energy is set to $-6.0$ Ry, and $\mathrm{R_{MT}K_{MAX}}$ is chosen to be 7. For the self-consistent field (SCF) calculation, 7 $\times$ 8 $\times$ 7 $k$-mesh of $\mathrm{SrAs_3}$ is used.

\begin{figure}[h]
\includegraphics[scale=1]{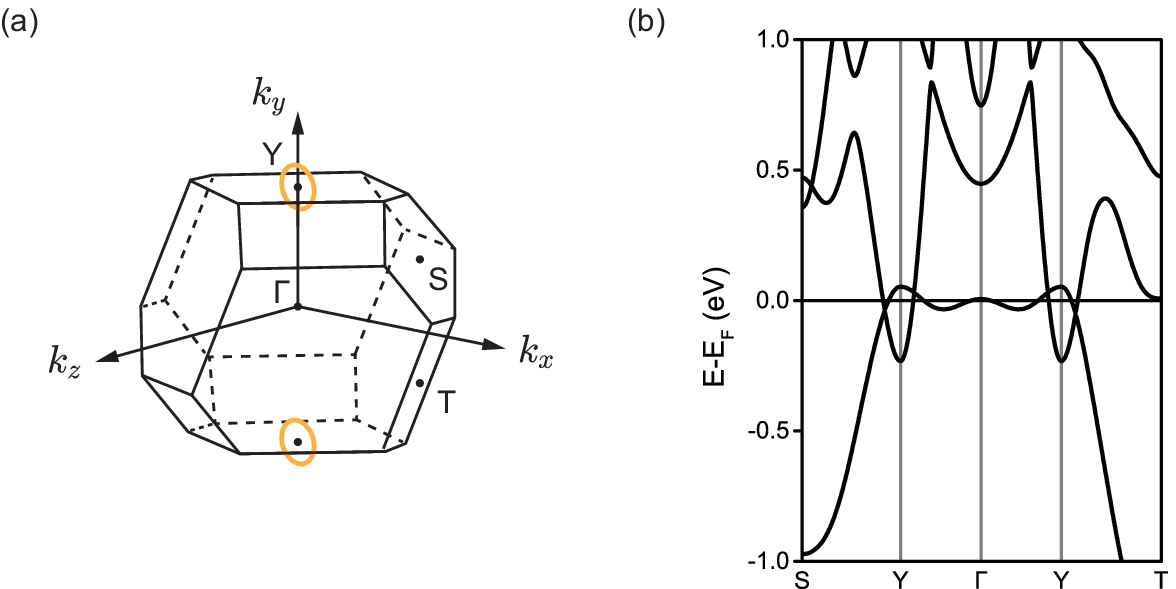}
\captionsetup{justification=raggedright,singlelinecheck=false}
\caption{Band structures of $\mathrm{SrAs_3}$ from first-principles calculations.}
\label{fig:sm:band}
\end{figure}

In Fig. \ref{fig:sm:band}, the band structure of $\mathrm{SrAs_3}$ is plotted along the high-symmetry points of the Brillouin zone, revealing a simple electronic structure with only a single nodal ring crossing the Fermi level near the $\mathrm{Y}$ point. This suggests that $\mathrm{SrAs_3}$ is an ideal candidate for studying nodal-ring semimetals.

\subsection{\uppercase\expandafter{\romannumeral3}-2. Theoretical phonon frequency calculation}

\begin{table}[h]
\centering
\caption{\label{table:sm:phonon}Calculated phonon frequencies}
\setlength{\tabcolsep}{15pt}
\label{t4}
\begin{tabular}{c|cc|cc}
\noalign{\smallskip}\noalign{\smallskip}\hline\hline
\multirow{2}{*}{(cm$^{-1}$)} & \multicolumn{2}{c|}{Phonon 1} & \multicolumn{2}{c|}{Phonon 2}\\
\cline{2-5}
      & $\omega_{calc}$ & Mode & $\omega_{calc}$ & Mode\\
\hline
 $E \parallel k_x$ & 98.74 & $B_u$ & 154.44 & $B_u$\\
 $E \parallel k_z$ & 138.76 & $A_u$ & 219.82 & $A_u$\\
\hline
\hline
\end{tabular}
\end{table}

First-principles calculation of phonon modes in $\mathrm{SrAs_3}$ is performed within the density functional perturbation theory \cite{togo2015first} routine using Vienna ab-initio simulation package \cite{PhysRevB.54.11169}.
After obtaining the dynamical matrix of the relaxed structure, post-processing is performed with Phonopy code to obtain phonon frequencies and modes at the zone center.
Phonon peaks shown in the IR conductivity measurement are matched with calculated IR active phonon frequencies considering group theory analysis.

\subsection{\uppercase\expandafter{\romannumeral3}-3. Model Hamiltonian}
The low-energy electronic structure of $\mathrm{SrAs_3}$ near the Y point can be described by the following minimal model Hamiltonian \cite{PhysRevB.95.045136, PhysRevLett.118.176402}:

\begin{widetext}
\begin{align} \label{eq:sm:ham}
H(\bm{k}) &= f_0(\bm{k})\sigma_0 s_0 + f_1(\bm{k})\sigma_1 s_0 + f_2(\bm{k})\sigma_2 s_0 + \Delta_{\rm SOC}\sigma_3 s_3, \\
f_0(\bm{k}) &= a_0 + a_x k_x^2 + a_{xy} k_x k_y + a_y k_y^2 + a_z k_z^2, \\
f_1(\bm{k}) &= b_0 + b_x k_x^2 + b_{xy} k_x k_y + b_y k_y^2 + b_z k_z^2, \label{eq:sm:ham_f1} \\
f_2(\bm{k}) &= \hbar v_z k_z,
\end{align}
\end{widetext}
where $\bm{k}$ is the wave vector measured from the Y point, and $\bm{\sigma}$ and $\bm{s}$ are Pauli matrices acting on the pseudospin and spin degrees of freedom, respectively, and $\Delta_{\rm SOC}$ is the strength of the spin-orbit coupling, which, as a leading approximation, we take as constant. Here $a_i$, $b_i$ ($i=0$, $x$, $xy$, $y$, $z$), and $v_z$ are material-dependent parameters determined from first-principles calculations and the experiments (see Sec. II-3)

The eigenvalues of the Hamiltonian are given by
\begin{equation}
E_{\pm}(\bm{k}) = f_0(\bm{k}) \pm \sqrt{|f_1(\bm{k})|^2 + |f_2(\bm{k})|^2 + \Delta_{\rm SOC}^2},
\end{equation}
where the conduction and valence bands are each doubly degenerate, and $f_0(\bm{k})$ is an energy tilt term. The band structure exhibits a gapped nodal ring defined by $f_1(\bm{k})=0$ and $f_2(\bm{k})=0$, that is,
\begin{equation} \label{eq:sm:nodal ring equation}
    b_x k_x^2 + b_{xy} k_x k_y + b_y k_y^2 = |b_0|, \ k_z=0.
\end{equation}
Note that this is an ellipse in the $k_z=0$ plane rotated by an angle $\phi_0 = 23.5^{\circ}$ clockwise with respect to the $k_x$ axis, and the lengths of the semi-major and semi-minor axes are given by $k_l = 0.073 \ \mathrm{\AA^{-1}}$ and $k_s = 0.063 \ \mathrm{\AA^{-1}}$, respectively, as shown in Fig. 1(d) in the main text with the average radius $k_0=\sqrt{k_l k_s}=0.068 \ \mathrm{\AA^{-1}}$ and $k_l/k_s = 1.16$.

\subsection{\uppercase\expandafter{\romannumeral3}-4. Determination of the model parameters} \label{sec:sm:params}
Overall, the band structure of $\mathrm{SrAs_3}$ calculated using the mBJ functional well describes the nodal-ring semimetal feature. However, there is a discrepancy in the optical conductivity between experimental results and calculations, such as the frequency where $\sigma_{zz}$ ($\sigma_{xx}$) starts to decrease (increase) and the position of the SOC-induced peak. These are related to the band overlap energy $2\varepsilon_0$ at the Y point and the strength of the SOC $\Delta_{\mathrm{SOC}}$, respectively. Moreover, the choice of the exchange-correlation functionals has a significant impact on the band overlap energy $2\varepsilon_0$, as the low-energy band structure is extremely sensitive to the exchange-correlation functional used in DFT calculations \cite{PhysRevB.95.045136,Kim2022}. 

Despite these limitations, the nodal-ring feature and the energy dispersion near the nodal point are retained, regardless of the choice of the exchange-correlation functional, except for their relative position. Therefore, we extract $2\varepsilon_0$ and $\Delta_{\mathrm{SOC}}$ from the experiment and use them as the model parameters. The other parameters in the model Hamiltonian in Eq. (\ref{eq:sm:ham}) are chosen to fit the band structure obtained from first-principles calculations.

\begin{table}[h]
\renewcommand{\arraystretch}{1.0}
\begin{tabular*}{0.9\linewidth}{@{\extracolsep{\fill}} cccccc }
\hline \hline
$a_{0}$ & $a_{x}$ & $a_{xy}$ & $a_{y}$ & $a_{z}$ & $\Delta_{\mathrm{SOC}}$ \\
 ($\mathrm{eV}$) & ($\mathrm{eV \cdot \AA^{2}}$) & ($\mathrm{eV \cdot \AA^{2}}$) & ($\mathrm{eV \cdot \AA^{2}}$) & ($\mathrm{eV \cdot \AA^{2}}$)  &  (eV) \\
  -0.0448 & 2.744 & -0.319 & 12.736  & -7.230 & 0.015 \\ \hline 
$m_{0}$ & $m_{x}$ & $m_{xy}$ & $m_{y}$ & $m_{z}$ & $\hbar v_z$ \\
 ($\mathrm{eV}$) & ($\mathrm{eV \cdot \AA^{2}}$) & ($\mathrm{eV \cdot \AA^{2}}$) & ($\mathrm{eV \cdot \AA^{2}}$) & ($\mathrm{eV \cdot \AA^{2}}$)  &  ($\mathrm{eV \cdot \AA^{1}}$) \\
  0.0645 & -12.849 & -3.091 & -15.741 & 9.215 & 2.292 \\ \hline \hline
\end{tabular*}
\caption{The parameters for the Hamiltonian in Eq. (\ref{eq:sm:ham}).}
\end{table}

To calculate the optical conductivity of $\mathrm{SrAs_3}$, as shown in Fig. 3(b) in the main text, we need to set the Fermi energy. According to field-dependent Hall resistivity and quantum oscillation measurements, the hole density $n_h$ is one or two orders of magnitude higher than the electron density $n_e$ \cite{Kim2022}. Using this information, we can obtain the corresponding Fermi energy $E_{\mathrm{F}}=-15$ meV through the density calculation based on the Hamiltonian in Eq. (\ref{eq:sm:ham}). The negative value of the Fermi energy implies that the sample is hole-doped, as the maximum energy of the valence band is set to zero.

\subsection{\uppercase\expandafter{\romannumeral3}-5. Methods for calculations of optical conductivity}
The optical conductivity is calculated within the Kubo formalism \cite{mahan2000many}:
\begin{align} \label{eq:sm:kubo}
    \sigma_{ij}(\omega) &= -\frac{ie^2}{\hbar} \sum_{s, s'} \int \frac{d^dk}{(2\pi)^d}\frac{f_{s,\bm{k}} - f_{s',\bm{k}}}{\varepsilon_{s, \bm{k}} - \varepsilon_{s', \bm{k}}} \frac{M_i^{ss'}(\bm{k}) M_j^{s's}(\bm{k})}{\hbar\omega+\varepsilon_{s, \bm{k}} - \varepsilon_{s', \bm{k}} + i\gamma},
\end{align}
where $i, j = x, y, z$ are spatial directions, $\varepsilon_{s, \bm{k}}$ and $f_{s, \bm{k}} = 1/[1+e^{(\varepsilon_{s, \bm{k}}-\mu)/(k_\mathrm{B} T)}]$ are the eigenenergy and the Fermi-Dirac distribution function for the band index $s$ and wave vector $\bm{k}$, respectively, $\mu$ is the chemical potential, and $M_i^{ss'}(\bm{k})=\braket{s,\bm{k}|\hbar\hat{v}_i|s',\bm{k}'}$ with the velocity operator $\hat{v}_i$ obtained from the relation $\hat{v}_i = \frac{1}{\hbar}\frac{\partial \hat{H}}{\partial k_i}$. $\gamma$ is a phenomenological broadening term proportional to the inverse lifetime which takes $\gamma \rightarrow 0^{+}$ for the clean limit. For numerical calculations, we set $\gamma=1$ meV. From now on, we only consider the real part of the longitudinal optical conductivity.

\section{\uppercase\expandafter{\romannumeral4}. Low-frequency asymptotic forms of the optical conductivity}
\subsection{\uppercase\expandafter{\romannumeral4}-1. Low-energy effective Hamiltonian}
Interestingly, the Hamiltonian in Eq. (\ref{eq:sm:ham}) can be rewritten as a form of a collection of gapped anisotropic graphene sheets at low energies, as we show explicitly below.
We introduce the polar coordinates $\bm{k} = (k_{\rho} \cos{\phi}, k_{\rho}\sin{\phi})$, which transforms $f_1(\bm{k})$ into the following form:
\begin{align}
    f_1(\bm{k}) &= \frac{\hbar^2}{2m_0}\left[
    \Phi(\phi)^2 k_{\rho}^2 - k_0^2 \right] + b_z k_z^2,
\end{align}
where $k_0 = \sqrt{k_l k_s}$ is the characteristic scale of the nodal ring size, which is nothing but the geometric mean of the lengths of the semi-major and semi-minor axes of the ellipse, and $m_0$ is the effective mass defined from the relation $b_0=\frac{\hbar^2 k_0^2}{2m_0}$. Here, $\Phi(\phi)$ describes the elliptical nature of the nodal ring defined by
\begin{align}
    \Phi(\phi) &= \sqrt{\frac{k_s}{k_l}\cos^2{(\phi+\phi_0)} + \frac{k_l}{k_s}\sin^2{(\phi+\phi_0)}},
\end{align}
which reduces to 1 for the circular case $(k_l = k_s)$. Note that the nodal ring in polar coordinates is given by $k_{\rho}(\phi) = k_0/\Phi(\phi)$. Near the nodal ring, by keeping only linear terms in $\bm{k}$, we can approximate Hamiltonian in Eq. (\ref{eq:sm:ham}) as 
\begin{align} \label{eq:sm:ham_gr}
    H(\bm{k}) = \Delta_{\mathrm{tilt}}(\bm{k})\sigma_0 s_0 + \hbar v_{\rho}(\phi) \delta k_{\rho} \sigma_1 s_0 + \hbar v_z k_z \sigma_2 s_0 + \Delta_{\mathrm {SOC}} \sigma_3 s_3,
\end{align}
where $v_{\rho}(\phi)$ is the angle-dependent velocity given by
\begin{equation} \label{eq:sm:velocity}
    v_{\rho}(\phi) =  \frac{\hbar k_0}{m_0} \Phi(\phi),
\end{equation}
and $\delta k_{\rho}$ is a wave vector measured from the nodal ring. Since the approximate Hamiltonian in Eq. (\ref{eq:sm:ham_gr}) contains only diagonal components $s_0$ and $s_3$ in spin space, it can be considered as two decoupled copies of $H_+$ and $H_-$ given by
\begin{align} \label{eq:sm:ham_gr_2by2}
    H_{\pm}(\bm{k}) = \Delta_{\mathrm{tilt}}(\bm{k})\sigma_0 + \hbar v_{\rho}(\phi) \delta k_{\rho} \sigma_1 + \hbar v_z k_z \sigma_2 \pm \Delta_{\mathrm {SOC}} \sigma_3.
\end{align}
Note that $H_+$ and $H_-$ have the same energy dispersion, leading to the same optical conductivity. This means that the total optical conductivity is $g=2$ times of the optical conductivity calculated from $H_+$ (or $H_-$), where $g$ denotes the band degeneracy factor. From now on, we will focus on $H_+$.

The Hamiltonian in Eq. (\ref{eq:sm:ham_gr_2by2}) takes the same form as that of graphene with different velocities ($v_{\rho}$ and $v_z$), an energy gap (2$\Delta_{\mathrm{SOC}}$) and an energy tilt ($\Delta_{\mathrm{tilt}}$).
We emphasize that the graphene approximation is valid only at low energies near the nodal ring where the phase space for allowed interband transitions for a given frequency forms a toroidal shape enclosing the nodal ring \cite{PhysRevLett.119.147402}.

\subsection{\uppercase\expandafter{\romannumeral4}-2. Optical conductivity as a sum of gapped anisotropic graphene sheets}

\begin{figure}[h]
\includegraphics[scale=1.0]{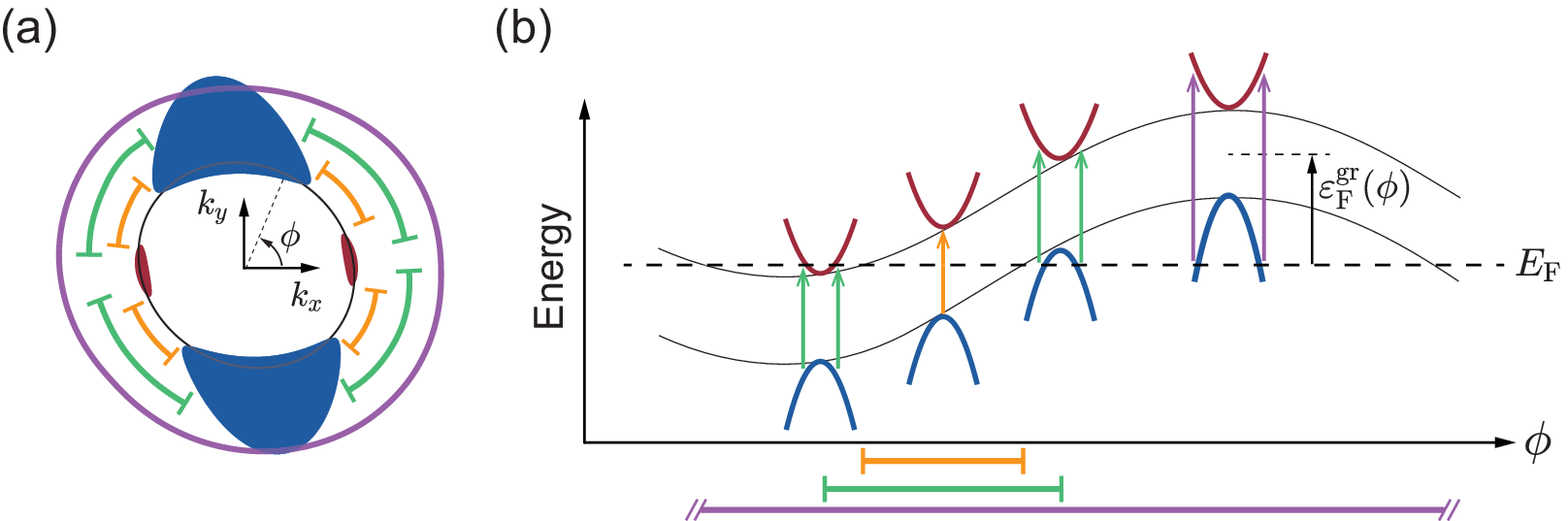}
\captionsetup{justification=raggedright,singlelinecheck=false}
\caption{(a) Schematic illustration of the Fermi surface in the $k_z=0$ plane, with the electron (hole) pocket indicated in red (blue) and the nodal ring depicted by the black solid line. (b) The corresponding band structures of graphene sheets are plotted for a selected value of $\phi$ with the conduction (valence) bands colored red (blue). The black curves represent the conduction band minimum and valence band maximum and the black dashed line represents the Fermi energy $E_{\mathrm{F}}$. The orange, green, purple arrows indicate the onset energies of the interband transitions at different values of $\phi$. The allowed ranges of angles for the corresponding interband transitions are presented with the same color in both panels (a) and (b).}
\label{fig:sm:sfig_th1}
\end{figure}

At low frequencies, the interband contribution to the optical conductivity can be obtained by summing all the contributions from each gapped anisotropic graphene sheet along the nodal ring (see Eq. (3) in the main text). The optical conductivity of a single Dirac cone in a gapped anisotropic graphene sheet [Eq. (\ref{eq:sm:ham_gr_2by2})] is given by \cite{doi:10.1073/pnas.1809631115}
\begin{align} \label{eq:sm:opcd_gr}
    \sigma_{ii}^{\mathrm{gr}}(\omega, \phi) &= \frac{e^2}{16\hbar} \frac{v_{i}(\phi)^2}{v_{\rho}(\phi) v_z} \left[1+\left(\frac{2\Delta_{\mathrm{SOC}}}{\hbar\omega}\right)^2\right] \Theta(\hbar\omega - 2 \mathrm{max}[\Delta_{\mathrm{SOC}}, |\varepsilon_{\mathrm{F}}^{\mathrm{gr}}(\phi)|]).
\end{align}
Here $\varepsilon_{\mathrm{F}}^{\mathrm{gr}}(\phi)$ is the angle-dependent Fermi energy of a graphene sheet measured from the middle of the gap. The angle dependence arises from the energy tilt term ($\Delta_{\mathrm{tilt}}$). Note that $\varepsilon_{\mathrm{F}}^{\mathrm{gr}}(\phi)$ differs from the Fermi energy of the material ($E_{\mathrm{F}}$), as shown in Fig. \ref{fig:sm:sfig_th1}. The optical conductivity vanishes in the low-frequency range due to the Pauli blocking for $\hbar\omega < 2|\varepsilon_{\mathrm{F}}^{\mathrm{gr}}(\phi)|$ or the SOC-induced optical gap for $\hbar\omega < 2\Delta_{\mathrm{SOC}}$. As the frequency increases, the optical conductivity approaches that without SOC, showing flat behavior.

Plugging Eq. (\ref{eq:sm:opcd_gr}) into Eq. (3) in the main text, we get
\begin{align}
    \label{eq:sm:opcd_int}
    \sigma_{ii}(\omega) &= g \frac{e^2}{16\hbar}
    \left[
    \int_0^{2\pi} \frac{d\phi}{2\pi} \frac{v_{i}(\phi)^2}{v_{\rho}(\phi) v_z} \mathcal{F}_{ii}(\phi) \Theta(\hbar\omega - 2 \mathrm{max}[\Delta_{\mathrm{SOC}}, |\varepsilon_{\mathrm{F}}^{\mathrm{gr}}(\phi)|])
    \right]
    \left[
    1+\left(\frac{2\Delta_{\mathrm{SOC}}}{\hbar\omega}\right)^2
    \right].
\end{align}
The step function $\Theta(x)$ term in Eq. (\ref{eq:sm:opcd_int}) inside the integral restricts the range of angle $\phi$ where the interband transitions are allowed for a given frequency $\omega$, making the calculation of the integral non-trivial. In order to understand the role of the Fermi energy $\varepsilon_{\mathrm F}^{\mathrm{gr}}(\phi)$ and $\Delta_{\mathrm{SOC}}$ in optical conductivity, we present plots of the Fermi surface and the corresponding band structures of graphene sheets for selected values of $\phi$ in Fig. \ref{fig:sm:sfig_th1}. In the current experiment, the Fermi energy $E_{\mathrm{F}}$ of $\mathrm{SrAs_3}$ lies within the SOC-induced gap for at least certain range of angle, as shown in Fig. \ref{fig:sm:sfig_th1}(b).

For $\hbar\omega < 2\Delta_{\mathrm{SOC}}$, we find that the $\Theta(x)$ term in Eq. (\ref{eq:sm:opcd_int}) vanishes for all $\phi$ as can be found in Fig. \ref{fig:sm:sfig_th1}(b), indicating that no interband transitions are allowed. At $\hbar\omega = 2\Delta_{\mathrm{SOC}}$, the step function becomes nonzero in a certain range of angles (orange lines in Fig. \ref{fig:sm:sfig_th1}) and the interband transitions set in. As the frequency increases, the range of angles for which interband transitions are allowed expands (green lines in Fig. \ref{fig:sm:sfig_th1}) and eventually reaches $0<\phi<2\pi$ (purple lines in Fig. \ref{fig:sm:sfig_th1}).

The important thing to note is that the threshold frequency for interband transitions is fixed at $\hbar\omega = 2\Delta_{\mathrm{SOC}}$ as long as the Fermi energy $E_{\mathrm{F}}$ lies within the SOC-induced gap for at least certain range of angle. As a result, the position of the SOC-induced optical peak is $\hbar\omega = 2\Delta_{\mathrm{SOC}}$, while the amplitude of the optical peak could vary depending on $E_{\mathrm{F}}$.

The angular integral in Eq. (\ref{eq:sm:opcd_int}) cannot be solved analytically. However, for frequencies above 2$\mathrm{max}[\varepsilon_{\mathrm{F}}^{\mathrm{gr}}(\phi)]$ where interband transitions are allowed for the entire range of angles ($0<\phi<2\pi$), we can analytically obtain the integral as follows:
\begin{align}
    \label{eq:sm:opcd_analytic}
    \sigma_{ii}(\omega) &=
        \sigma_{ii}^{\mathrm{flat}} 
        \left[
        1+\left(\frac{2\Delta_{\mathrm{SOC}}}{\hbar\omega}\right)^2
        \right] \Theta(\hbar\omega - 2 \Delta_{\mathrm{SOC}}),
\end{align}
where
\begin{align}
    \label{eq:sm:opcd_flat_int}
    \sigma_{ii}^{\mathrm{flat}} =
    g \frac{e^2 k_0}{16\hbar}
    \biggl(
    \int_0^{2\pi} \frac{d\phi}{2\pi} \frac{v_{i}(\phi)^2}{v_{\rho}(\phi) v_z} \mathcal{F}_{ii}(\phi) \biggr).
\end{align}
The term inside the square bracket in Eq. (\ref{eq:sm:opcd_analytic}) represents the effect of SOC and determines the amplitude of the SOC-induced optical peak. When interband transitions are allowed only within a limited range of angles, the square bracketed term is modified and can only be obtained numerically.

\subsection{\uppercase\expandafter{\romannumeral4}-3. Calculating the flat optical conductivity}
In this subsection, we present methods to calculate the flat conductivity $\sigma_{ii}^{\mathrm{flat}}$ analytically. Let us consider the following Hamiltonian describing the
circular nodal ring with radius $r_0$ and velocity $v$ for both the radial and $k_z$ directions:
\begin{equation} \label{eq:sm:ham_flat_circle}
    H_0(\bm{k}) = \hbar v\left[\sqrt{\left(\frac{k_x}{r_0}\right)^2 + \left(\frac{k_y}{r_0}\right)^2} - 1\right]\sigma_1 + \hbar v k_z \sigma_2,
\end{equation}
and denote the corresponding flat optical conductivity as $\sigma^{\mathrm{flat, 0}}_{ii}$. Then the corresponding optical conductivity is given by $\sigma^{\mathrm{flat, 0}}_{xx} = \sigma^{\mathrm{flat, 0}}_{yy} = \sigma^{\mathrm{flat, 0}}_{zz}/2 = r_0 e^2/(32\hbar)$ \cite{PhysRevLett.119.147402}. Next, consider the optical conductivity of the elliptical ring. After the following coordinate scaling, $k_x \rightarrow \frac{r_0}{k_l} k_x$, $k_y \rightarrow \frac{r_0}{k_s} k_y$, $k_z \rightarrow \frac{v_z}{v} k_z$, the Hamiltonian $H_0(\bm{k})$ in Eq. (\ref{eq:sm:ham_flat_circle}) is transformed into
\begin{equation} \label{eq:sm:ham_flat_ellipse}
    \tilde{H}(\bm{k}) = \hbar v\left[\sqrt{\left(\frac{k_x}{k_l}\right)^2 + \left(\frac{k_y}{k_s}\right)^2} - 1\right]\sigma_1 + \hbar v_z k_z \sigma_2,
\end{equation}
where the nodal ring is no longer a circle but an ellipse with the lengths of the semi-major and semi-minor axes $k_l$ and $k_s$, respectively, and the Fermi velocities along the in-plane and out-of-plane directions are $v$ and $v_z$, respectively.

Denoting $\tilde{\sigma}^{\mathrm{flat}}_{ii} = \tilde{\sigma}^{\mathrm{flat}}_{ii}(k_l,k_s,v/v_z)$ as the flat conductivity obtained from $\tilde{H}(\bm{k})$ in Eq. (\ref{eq:sm:ham_flat_ellipse}), we get the following scaling relation of the flat optical conductivity from the Kubo formula in Eq. (\ref{eq:sm:kubo}):
\begin{align}
    \tilde{\sigma}^{\mathrm{flat}}_{xx} &= \left(\frac{v}{v_z}\right)\left(\frac{k_s}{k_l}\frac{\sqrt{k_l k_s}}{r_0}\right)\sigma^{\mathrm{flat, 0}}_{xx}, \label{eq:sm:sigma_xx}\\
    \tilde{\sigma}^{\mathrm{flat}}_{yy} &= \left(\frac{v}{v_z}\right)\left(\frac{k_l}{k_s}\frac{\sqrt{k_l k_s}}{r_0}\right)\sigma^{\mathrm{flat, 0}}_{yy}, \\
    \tilde{\sigma}^{\mathrm{flat}}_{zz} &= \left(\frac{v_z}{v}\right)\left(\frac{\sqrt{k_l k_s}}{r_0}\right)\sigma^{\mathrm{flat, 0}}_{zz} \label{eq:sm:sigma_zz}.
\end{align}

Finally, we rotate the $k_x$ and $k_y$ axes by $\phi_0$ clockwise. The rotated Hamiltonian corresponds to the model Hamiltonian in Eq. (\ref{eq:sm:ham}). Then the flat optical conductivity matrix transforms to $\sigma_{1}^{\mathrm{flat}}(k_l, k_s, v/v_z, \phi_0) = R^{-1} \tilde{\sigma_{1}}^{\mathrm{flat}}(k_l, k_s, v/v_z) R$, where $R$ is the relevant rotation matrix. The longitudinal components are then given as
\begin{align}
\sigma^{\mathrm{flat}}_{xx}(k_l, k_s, v/v_z, \phi_0) &= \frac{1}{2} \left[\tilde{\sigma}^{\mathrm{flat}}_{xx} + \tilde{\sigma}^{\mathrm{flat}}_{yy} + \left(\tilde{\sigma}^{\mathrm{flat}}_{xx} - \tilde{\sigma}^{\mathrm{flat}}_{yy}\right)\cos2\phi_0 \right], \label{eq:sm:sigma_xx_rot} \\
\sigma^{\mathrm{flat}}_{yy}(k_l, k_s, v/v_z, \phi_0) &= \frac{1}{2} \left[\tilde{\sigma}^{\mathrm{flat}}_{xx} + \tilde{\sigma}^{\mathrm{flat}}_{yy} - \left(\tilde{\sigma}^{\mathrm{flat}}_{xx} - \tilde{\sigma}^{\mathrm{flat}}_{yy}\right)\cos2\phi_0 \right],\\
\sigma^{\mathrm{flat}}_{zz}(k_l, k_s, v/v_z, \phi_0) &= \tilde{\sigma}^{\mathrm{flat}}_{zz}. \label{eq:sm:sigma_zz_rot}
\end{align}
Plugging Eqs. (\ref{eq:sm:sigma_xx}-\ref{eq:sm:sigma_zz}) into Eqs. (\ref{eq:sm:sigma_xx_rot}-\ref{eq:sm:sigma_zz_rot}), we find the analytic form of the flat optical conductivity (Eq. (5) in the main text) given by
\begin{align}
\sigma^{\mathrm{flat}}_{xx}(k_l, k_s, v/v_z, \phi_0) &= \frac{1}{2}\frac{v}{v_z} \frac{\sqrt{k_l k_s}}{r_0}
\left[\frac{k_s}{k_l} + \frac{k_l}{k_s} + \left(\frac{k_s}{k_l} - \frac{k_l}{k_s}\right)\cos2\phi_0 \right] \sigma^{\mathrm{flat},0}_{xx} \nonumber \\
&= k_0 \frac{e^2}{16\hbar} \frac{g}{4} \frac{v}{v_z} \left[\frac{k_s}{k_l}+\frac{k_l}{k_s} + \left(\frac{k_s}{k_l}-\frac{k_l}{k_s}\right) \cos{2\phi_0} \right], \\
\sigma^{\mathrm{flat}}_{yy}(k_l, k_s, v/v_z, \phi_0) &= \frac{1}{2}\frac{v}{v_z} \frac{\sqrt{k_l k_s}}{r_0}
\left[\frac{k_s}{k_l} + \frac{k_l}{k_s} - \left(\frac{k_s}{k_l} - \frac{k_l}{k_s}\right)\cos2\phi_0 \right] \sigma^{\mathrm{flat},0}_{yy} \nonumber \\
&= k_0 \frac{e^2}{16\hbar} \frac{g}{4} \frac{v}{v_z} \left[\frac{k_s}{k_l}+\frac{k_l}{k_s} - \left(\frac{k_s}{k_l}-\frac{k_l}{k_s}\right) \cos{2\phi_0} \right], \\
\sigma^{\mathrm{flat}}_{zz}(k_l, k_s, v/v_z, \phi_0) &= \frac{v_z}{v}\frac{\sqrt{k_l k_s}}{r_0}\sigma^{\mathrm{flat},0}_{zz} \nonumber \\
&= k_0 \frac{e^2}{16\hbar} g \frac{v_z}{v},
\end{align}
where $k_0 = \sqrt{k_l k_s}$ and $g=2$ is the band degeneracy factor.

The analytic form of the flat conductivity can be directly calculated from the integral in Eq. (\ref{eq:sm:opcd_flat_int}) with the geometric factors $\mathcal{F}_{ii}(\phi)$ that describe the projection of the polarization direction on the graphene sheet and that of the graphene velocity on the current direction. The geometric factors can be obtained from the Kubo formula in Eq. (\ref{eq:sm:kubo}) and are given by
\begin{align}
    \mathcal{F}_{xx}(\phi) &= k_0 \cos^2{\left(\phi+\phi_0\right)} \left[\frac{k_s^4}{\left(\sqrt{(k_s\cos{\left(\phi+\phi_0\right)})^2+(k_l\sin{\left(\phi+\phi_0\right)})^2}\right)^{5}} \right], \\
    \mathcal{F}_{yy}(\phi) &= k_0 \sin^2{\left(\phi+\phi_0\right)} \left[\frac{k_l^4}{\left(\sqrt{(k_s\cos{\left(\phi+\phi_0\right)})^2+(k_l\sin{\left(\phi+\phi_0\right)})^2}\right)^{5}} \right], \\
    \mathcal{F}_{zz}(\phi) &=  \frac{1}{k_0}\sqrt{(k_s\cos{\left(\phi+\phi_0\right)})^2+(k_l\sin{\left(\phi+\phi_0\right)})^2}.
\end{align}
Note that for a circular nodal ring with $v_{\rho}=v_z$, $\mathcal{F}_{ii}(\phi)$ reduce to $\mathcal{F}_{xx}(\phi)=\cos^2{\left(\phi+\phi_0\right)}$, $\mathcal{F}_{yy}(\phi)=\sin^2{\left(\phi+\phi_0\right)}$, and $\mathcal{F}_{zz}(\phi)=1$, respectively.

\section {\uppercase\expandafter{\romannumeral5}. Temperature dependence of the chemical potential}

\begin{figure}[!h]
\includegraphics[width=0.5\columnwidth]{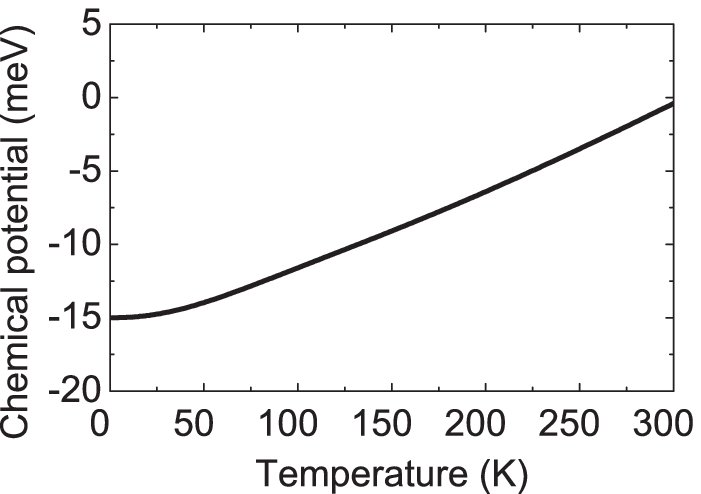}
\caption{Calculated chemical potential as a function of temperature.}
\label{fig:sm:chemical}
\end{figure}

The carrier density is given by
\begin{align} \label{eq:sm:density}
    n = \int_{-\infty}^{\infty} d\varepsilon \ D(\varepsilon) f(\varepsilon, \mu(T), T)
    = \int_{-\infty}^{\infty} d\varepsilon \ D(\varepsilon) f(\varepsilon, E_{\mathrm{F}}, 0)
\end{align}
where $D(\varepsilon)$ is the density of states, $f(\varepsilon, \mu, T)=1/[1+e^{(\varepsilon-\mu)/(k_\mathrm{B} T)}]$ is the Fermi-Dirac distribution function at the chemical potential $\mu$ for a given temperature $T$, and $E_{\mathrm{F}}=\mu(T=0)$ is the Fermi energy.
In calculating the carrier density, we use the model parameters in Table S3.
Since the number of carriers remains constant with respect to temperature changes, the variation of the chemical potential as a function of temperature can be given by solving Eq. (\ref{eq:sm:density}) for $\mu(T)$. Figure \ref{fig:sm:chemical} shows the numerical solution of the chemical potential as a function of temperature with the Fermi energy $E_{\mathrm{F}}=-15$ meV. As temperature increases, the chemical potential is shifted towards regions of low density of states.

\bibliographystyle{apsrev4-2}
\bibliography{sm}